\documentclass[twocolumn,aps,prx,footinbib,floatfix,nolongbibliography,superscriptaddress]{revtex4-2}
\usepackage[utf8]{inputenc}

\usepackage{amsmath,amssymb,bm,graphicx,siunitx}
\usepackage[colorlinks=true, linkcolor=blue, citecolor=blue, pdfencoding=auto]{hyperref}
\usepackage[dvipsnames]{xcolor}
\usepackage[export]{adjustbox}
\usepackage[normalem]{ulem} 

\renewcommand{\Re}{{\rm Re}\,}
\newcommand{\sign}{{\rm sign}\,}

\newcommand{\ket}[1]{|#1\rangle}

\newcommand{\braket}[2]{\langle #1|#2\rangle}
\newcommand{\braOket}[3]{\langle #1|#2|#3\rangle}

\usepackage{tabularray}
\usepackage{tikz}
\usetikzlibrary{decorations.markings,arrows,arrows.meta}

\newcommand{\rBZtwo}{
\begin{tikzpicture}[baseline]
\def\R{1.25}; \draw (-0.75*\R,-0.5*\R) -- (-0.75*\R,0.5*\R) -- (0.75*\R,0.5*\R) -- (0.75*\R,-0.5*\R) -- (-0.75*\R,-0.5*\R); \node[inner sep=1.5pt, circle, fill, color=black, label=below:{$\gamma$}] at (0,0) {}; \node[inner sep=1.5pt, circle, fill, color=BrickRed, label=above:{$m$}] at (0.75*\R,0.5*\R) {}; \node[inner sep=1.5pt, circle,fill, color=BrickRed, opacity=0.35] at (-0.75*\R,-0.5*\R) {}; \draw[BrickRed, -{Stealth[scale=1.35]}] (0.75*\R,0.5*\R) to[bend right=22] (-0.75*\R,-0.5*\R) node[below] {$V_{g_m}$}; \node[inner sep=1.5pt, circle, fill, color=PineGreen, label=right:{$x$}] at (0.75*\R,0) {}; \node[inner sep=1.5pt, circle, fill, color=PineGreen, opacity=0.35] at (-0.75*\R,0) {}; \draw[PineGreen, -{Stealth[scale=1.35]}] (0.75*\R,0) to[bend right=19] (-0.75*\R,0) node[left] {$V_{g_x}$}; \node[inner sep=1.5pt, circle, fill, color=MidnightBlue, label=above:{$y$}] at (0,0.5*\R) {}; \node[inner sep=1.5pt, circle, fill, color=MidnightBlue, opacity=0.35] at (0,-0.5*\R) {}; \draw[MidnightBlue, -{Stealth[scale=1.35]}] (0,0.5*\R) to[bend right=35] (0,-0.5*\R) node[below] {$V_{g_y}$};
\end{tikzpicture}
}

\newcommand{\rBZthree}{
\begin{tikzpicture}[baseline]
\def\R{0.8}; \draw (0:\R) \foreach \x in {60,120,...,360} {  -- (\x:\R) };
\node[inner sep=1.5pt, circle, fill, color=black, label=below:{$\gamma$}] at (0,0) {}; \node[inner sep=1.5pt, circle, fill, color=BrickRed, label=left:{$\kappa$}] at (120:\R) {}; \node[inner sep=1.5pt, circle,fill, color=BrickRed, opacity=0.35] at (-120:\R) {}; \draw[BrickRed, -{Stealth[scale=1.35]}] (120:\R) to (-120:\R) node[below left] {$V_{g_\kappa}$}; \node[inner sep=1.5pt, circle, fill, color=PineGreen, label=right:{$\kappa'$}] at (60:\R) {}; \node[inner sep=1.5pt, circle,fill, color=PineGreen, opacity=0.35] at (180:\R) {}; \draw[PineGreen, -{Stealth[scale=1.35]}] (60:\R) node[xshift=-0.1cm,yshift=0.35cm] {$V_{g_{\kappa'}}=V_{g_\kappa}^*$} to (180:\R) ;
\end{tikzpicture} 
}

\newcommand{\rBZfour}{
\begin{tikzpicture}[baseline]
\def\R{0.625}; \draw (-\R,-\R) -- (-\R,\R) -- (\R,\R) -- (\R,-\R) -- (-\R,-\R); \node[inner sep=1.5pt, circle, fill, color=black, label=below:{$\gamma$}] at (0,0) {}; \node[inner sep=1.5pt, circle, fill, color=BrickRed, label=right:{$m$}] at (\R,\R) {}; \node[inner sep=1.5pt, circle,fill, color=BrickRed, opacity=0.35] at (-\R,\R) {}; \draw[BrickRed, -{Stealth[scale=1.35]}] (\R,\R) to[bend left=22] node[midway,label=above:{$V_{g_m}=V_{g_x}$}] {} (-\R,\R); \node[inner sep=1.5pt, circle, fill, color=PineGreen, label=right:{$x$}] at (\R,0) {}; \node[inner sep=1.5pt, circle, fill, color=PineGreen, opacity=0.35] at (-\R,0) {}; \draw[PineGreen, -{Stealth[scale=1.35]}] (\R,0) to[bend right=22] (-\R,0) node[left] {$V_{g_x}$};
\end{tikzpicture}
}

\newcommand{\rBZsix}{
\begin{tikzpicture}[baseline]
\def\R{0.8}; \draw (0:\R) \foreach \x in {60,120,...,360} {  -- (\x:\R) };
\node[inner sep=1.5pt, circle, fill, color=black, label=below:{$\gamma$}] at (0,0) {}; \node[inner sep=1.5pt, circle, fill, color=BrickRed, label=above:{$\kappa$}] at (120:\R) {}; \node[inner sep=1.5pt, circle,fill, color=BrickRed, opacity=0.35] at (-120:\R) {}; \draw[BrickRed, -{Stealth[scale=1.35]}] (120:\R) to (-120:\R) node[below left] {$V_{g_\kappa}$}; \node[inner sep=1.5pt, circle, fill, color=PineGreen, label=above:{$m$}] at (90:0.866*\R) {}; \node[inner sep=1.5pt, circle,fill, color=PineGreen, opacity=0.35] at (-90:0.866*\R) {}; \draw[PineGreen, -{Stealth[scale=1.35]}] (90:0.866*\R) to[bend left=30]  (-90:0.866*\R) node[xshift=0.5cm,yshift=-0.25cm] {$V_{g_m}=V_{g_\kappa}$};
\end{tikzpicture} 
}

\newcommand{\loopgamma}{
\begin{tikzpicture}[baseline]
\def\R{0.625}; \draw (-\R,-0.5*\R) -- (-\R,0.5*\R) -- (\R,0.5*\R) -- (\R,-0.5*\R) -- (-\R,-0.5*\R); \node[inner sep=1.5pt, circle, fill, color=black, label=right:{$\gamma$}] at (0,0) (origin) {}; \coordinate (top) at (-0.5*\R,0.25*\R); \node at (0,-0.6*\R) {}; \node at (0,0.6*\R) {}; \draw[fill,color=BrickRed,opacity=0.2] (origin) to[out=135,in=45,looseness=0.6] (top) to[out=-135,in=160,looseness=0.6] (origin); \draw[BrickRed,-{Stealth[scale=1.35]}] (origin) to[out=135,in=45,looseness=0.6] (top) to[out=-135,in=160,looseness=0.6] (origin); 
\end{tikzpicture}
}

\newcommand{\looppequaltwo}{
\begin{tikzpicture}[baseline]
\def\R{0.625}; \draw (-\R,-0.5*\R) -- (-\R,0.5*\R) -- (\R,0.5*\R) -- (\R,-0.5*\R) -- (-\R,-0.5*\R); \node[inner sep=1.5pt, circle, fill, color=black, label=right:{$m$}] at (\R,0.5*\R) (origin) {}; \node[inner sep=1.5pt, circle, fill, color=black, opacity=0.2] at (-\R,-0.5*\R) {}; \coordinate (top) at (-\R,-0.5*\R); \node at (0,-0.6*\R) {}; \node at (-1.7*\R,0) {}; \node at (0,0.6*\R) {}; \draw[fill,color=BrickRed,opacity=0.2] (origin) to[out=-170,in=50,looseness=0.6] (top) to[out=10,in=-130,looseness=0.6] (origin); \draw[BrickRed,-{Stealth[scale=1.35]}] (origin.center) to[out=-170,in=50,looseness=0.6] (top.center);  \draw[BrickRed,-{Stealth[scale=1.35]}] (top.center) to[out=10,in=-130,looseness=0.6] (origin.center); 
\end{tikzpicture}
}

\newcommand{\looppequalthree}{
\begin{tikzpicture}[baseline]
\def\R{0.8}; \draw (0:\R) \foreach \x in {60,120,...,360} {  -- (\x:\R) }; \node at (-1.2*\R,0) {}; \node at (0,-0.9*\R) {}; \node[inner sep=1.5pt, circle, fill, color=black, label=left:{$\kappa$}] at (120:\R) (origin) {}; \node[inner sep=1.5pt, circle, fill, color=black, opacity=0.2] at (240:\R) (R1) {}; \node[inner sep=1.5pt, circle, fill, color=black, opacity=0.2] at (0:\R) (R2) {}; \draw[BrickRed,fill,opacity=0.2] (origin.center) to[bend right=10] (R1.center) to[bend right=10] (R2.center) to[bend right=10]  (origin.center); \draw[BrickRed,-{Stealth[scale=1.35]}] (origin.center) to[bend right=10] (R1.center); \draw[BrickRed,-{Stealth[scale=1.35]}] (R1.center) to[bend right=10] (R2.center); \draw[BrickRed,-{Stealth[scale=1.35]}] (R2.center) to[bend right=10] (origin.center);
\end{tikzpicture}
}

\newcommand{\looppequalfour}{
\begin{tikzpicture}[baseline,decoration={markings,mark=at position 0.5 with {\arrow{{Stealth[scale=1.35]}}}}]
\def\R{0.625}; \draw (-\R,-\R) -- (-\R,\R) -- (\R,\R) -- (\R,-\R) -- (-\R,-\R); \node[inner sep=1.5pt, circle, fill, color=black, label=right:{$m$}] at (\R,\R) (origin) {}; \node[inner sep=1.5pt, circle, fill, color=black, opacity=0.2] at (-\R,\R) (R1) {}; \node[inner sep=1.5pt, circle, fill, color=black, opacity=0.2] at (-\R,-\R) (R2) {}; \node[inner sep=1.5pt, circle, fill, color=black, opacity=0.2] at (\R,-\R) (R3) {}; \node at (0,-1.1*\R) {}; \node at (0,1.1*\R) {};  \node at (-1.7*\R,0) {}; \draw[BrickRed,fill,opacity=0.2] (origin.center) to[bend right=10] (R1.center) to[bend right=10] (R2.center) to[bend right=10] (R3.center) to[bend right=10] (origin.center); \draw[BrickRed,-{Stealth[scale=1.35]}] (origin.center) to[bend right=10] (R1.center); \draw[BrickRed,-{Stealth[scale=1.35]}] (R1.center) to[bend right=10] (R2.center); \draw[BrickRed,-{Stealth[scale=1.35]}] (R2.center) to[bend right=10] (R3.center); \draw[BrickRed,-{Stealth[scale=1.35]}] (R3.center) to[bend right=10] (origin.center);
\end{tikzpicture}
}

\begin{document}
\title{
Efficient prediction of superlattice and anomalous miniband \\ topology from  quantum geometry 
}

\author{Valentin Cr\'epel}
\affiliation{Center for Computational Quantum Physics, Flatiron Institute, New York, New York 10010, USA}
\author{Jennifer Cano}
\affiliation{Department of Physics and Astronomy, Stony Brook University, Stony Brook, New York 11794, USA}
\affiliation{Center for Computational Quantum Physics, Flatiron Institute, New York, New York 10010, USA}

\begin{abstract}
Two dimensional materials subject to long-wavelength modulations have emerged as novel platforms to study topological and correlated quantum phases. 
In this article, we develop a versatile and computationally inexpensive method to predict the topological properties of materials subjected to a superlattice potential by combining degenerate perturbation theory with the method of symmetry indicators. 
In the absence of electronic interactions, our analysis provides a systematic rule to find the Chern number of the superlattice-induced miniband starting from the harmonics of the applied potential and a few material-specific coefficients. 
Our method also applies to anomalous (interaction-generated) bands, for which we derive an efficient algorithm to determine all Chern numbers compatible with a self-consistent solution to the Hartree-Fock equations. 
Our approach gives a microscopic understanding of the quantum anomalous Hall insulators recently observed in rhombohedral graphene multilayers. 
\end{abstract}

\maketitle

\section{Introduction}

The advent of moir\'e and, more generally, superlattice materials, has ushered in a new era in condensed matter physics, providing fertile ground for the exploration and design of exotic quantum phases~\cite{andrei2021marvels,mak2022semiconductor,ghorashi2023topological,nuckolls2024microscopic}. 
These systems, characterized by periodic modulations on the tens of nanometer scale, exhibit rich and tunable electronic properties due to the interplay between the low-energy physics of their constituents and the geometric properties of the long-wavelength modulation. 
Reverse-engineering this rich phenomenology to identify the optimal combination of materials and long-wavelength modulation to observe a specific phase of interest has, however, proven to be a very challenging task. 
Indeed, understanding the phase diagram of a particular material platform has traditionally relied on the computation of the full superlattice band structure -- a process often computationally intense and analytically opaque~\cite{uchida2014atomic,jung2014ab,carr2017twistronics,massatt2017electronic,zhang2021electronic,zhang2022n} --  which is then used to derive effective models~\cite{bistritzer2011moire,wu2019topological,Devakul_2021,crepel2024chiral,vafek2023continuum,kong2024modeling} that are finally studied using (costly) state-of-the-art numerical methods.

In this work, we propose an approximate but quick and versatile method for determining the topological phase diagram of superlattice materials using a few material-specific coefficients, circumventing the need for exhaustive band structure computations. 
More precisely, we rely on degenerate perturbation theory to analytically compute eigenvalues of crystalline rotation symmetry operators acting on the superlattice-induced minibands' eigenvectors at high-symmetry points of the reduced Brillouin zone determined by the superlattice potential. 
Our perturbative result requires only the knowledge of the superlattice geometry and a small number of form factors from the original material.
The Chern number of the minibands can be extracted from these eigenvalues using the symmetry indicators from Ref.~\cite{fang2012bulk}. 
This analytic topological criterion is correct as long as the gaps opened at high-symmetry points for perturbative values of the superlattice amplitude do not close as the latter is tuned to its microscopically realistic value. 
While we primarily focus on superlattice materials, which are elegantly captured by a single topological phase diagram for each symmetry class
(Tab.~\ref{tab_fullsummary}), the perturbative method 
that we introduce is highly versatile: we successfully apply it to continuum models for twisted transition metal dichalcogenides and twisted bilayer graphene in App.~\ref{app_homobilayers}. 

\begin{table*}
\centering
\begin{tblr}{colspec = {|X[c,m,3cm]||X[c,m,3.2cm]|X[c,m,3.1cm]|X[c,m,3.1cm]|X[c,m,3.1cm]|}}
\hline
Symmetry group $\mathcal{C}_n$ & $n=2$ & $n=3$ & $n=4$ & $n=6$ 
\\ \hline
Reduced BZ & \rBZtwo & \rBZthree & \rBZfour & \rBZsix 
\\ \hline \hline
Resonant harmonics determining $C_n$ in the long-period limit & $V_{g_x}, V_{g_y}, V_{g_m} \in \mathbb{R}$ & $V_{g_\kappa} = |V| e^{i \theta} \in \mathbb{C}$ & $V_{g_x} \in \mathbb{R}$ & $V_{g_\kappa} \in \mathbb{R}$ 
\\ \hline
Chern number $C_n$ in the long-period limit & $\Theta ( - V_{g_x} V_{g_y} V_{g_m})$ & ${\displaystyle\sum_{\epsilon = \pm 1}} \left\lfloor \frac{2\pi+ 6 \epsilon \theta - \phi_{\rm B} }{4\pi}  \right\rfloor$ & $\left\lfloor \frac{\pi - \phi_{\rm B}}{2\pi}  \right\rfloor$ & $\left. \begin{array}{c} 1 - v  + 2 \left\lfloor \frac{2 \pi v -  \phi_{\rm B}}{4\pi}  \right\rfloor \\ \text{with:} \; v = \Theta(V_{g_\kappa}) \end{array} \right.$
\\ \hline\hline 
Form factors for the general case & $\left. \begin{array}{c} M_q = \braOket{q}{U_2}{q} \\ q\in \{ \gamma , x, y, m \} \end{array} \right.$ & $\left. \begin{array}{c} M_q = \braOket{q}{U_3}{q} \\ q\in \{ \gamma , \kappa, \kappa' \} \end{array} \right.$ & $\left. \begin{array}{c} M_\gamma = \braOket{\gamma}{U_4}{\gamma}  \\ M_{m} = \braOket{m}{U_4}{m} \\ M_x = \braOket{x}{U_4^2}{x} \end{array} \right.$ & $\left. \begin{array}{c} M_\gamma = \braOket{\gamma}{U_6}{\gamma} \\ M_{m} = \braOket{m}{U_6^3}{m} \\ M_\kappa = \braOket{\kappa}{U_6^2}{\kappa} \end{array} \right.$ 
\\ \hline
Chern number $C_n$ in the general case \newline \vspace{5pt} \newline as a function of the loop phases $\Phi_q^{\circlearrowleft} = p(q) \arg \frac{V_{g_q} M_q^*}{M_\gamma^*} $ & \includegraphics[width=3.1cm,valign=m]{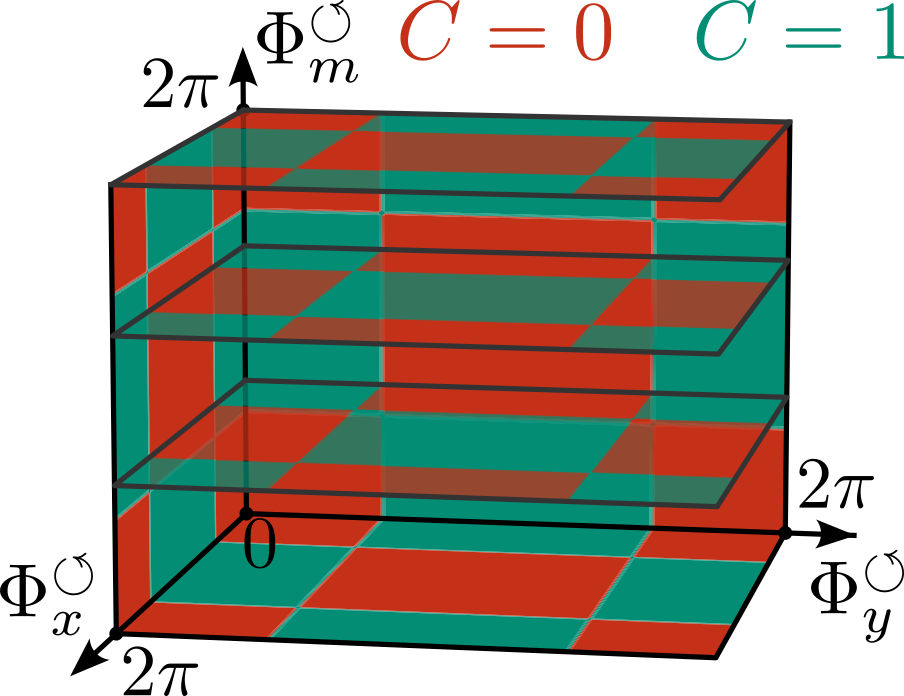} & \includegraphics[width=3.1cm,valign=m]{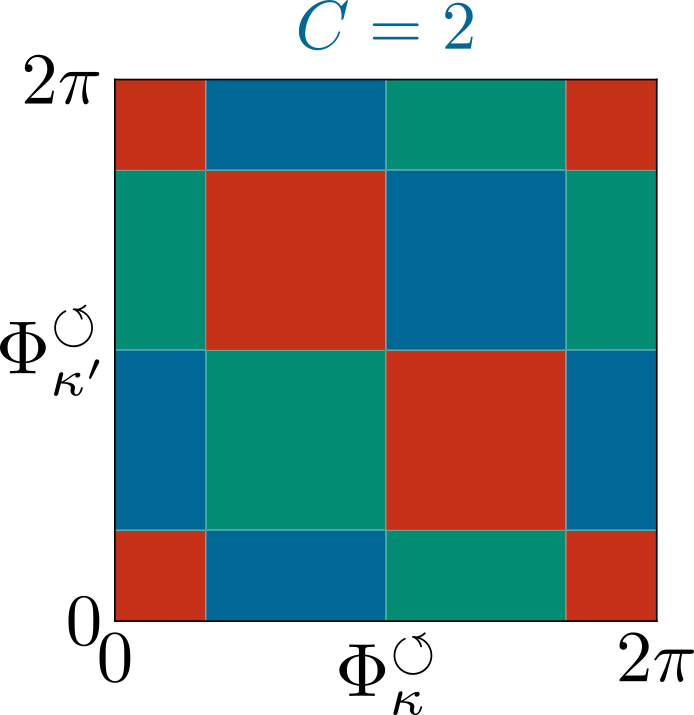} & \includegraphics[width=3.1cm,valign=m]{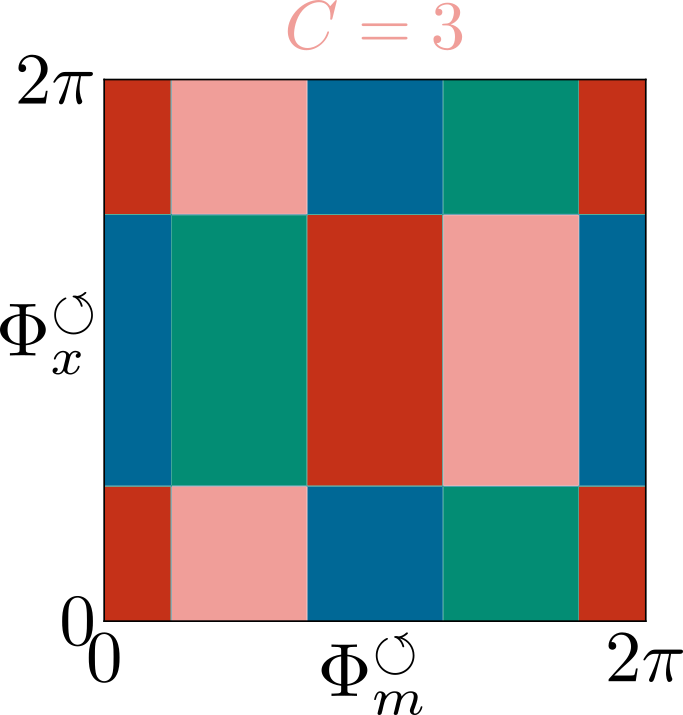} & \includegraphics[width=3.1cm,valign=m]{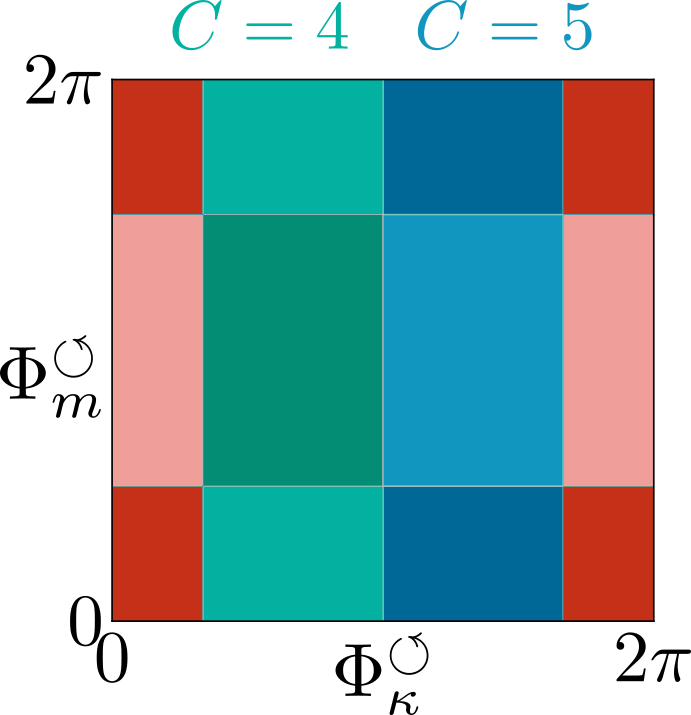}
\\ \hline
\end{tblr}
\caption{Topology of the lowest superlattice-induced miniband determined by degenerate perturbation theory using symmetry indicators, in the absence of interactions. 
The first row specifies the $\mathcal{C}_n$-rotational symmetry of the superlattice; $n=2,3,4$, or $6$. 
The second row indicates the high-symmetry points of the reduced Brillouin zone (rBZ) fixed by the superlattice, and the harmonics of the potential (third row) that resonantly couple these high-symmetry points to their degenerate rotated images. 
Row four gives the Chern number $C_n = C \, ({\rm mod} \, n)$ of the lowest superlattice-induced miniband in the long-period limit, where the rBZ is small compared to the scale on which the quantum geometry of the original material varies. 
In this limit, the Chern number only depends on a finite set of superlattice harmonics (real for even $n$, see row three) and on the Berry flux $\phi_{\rm B}$ of the original material enclosed in the rBZ. 
Here $\lfloor x \rfloor$ denotes the floor function, $\Theta$ the Heavyside function, and $\theta$ the phase of the only relevant complex harmonic (when $n=3$). 
Away from this limit, $C_n$ can still be analytically determined within our framework by replacing the Berry flux $\phi_{\rm B}$ by a small number of material-specific form factors $M_q$, which are gathered in row five; $U_p$ denotes the matrix representation of the $\frac{2\pi}{p}$-rotation, 
which acts on a Bloch eigenvector $\ket{q}$ with quasi-momentum $q$. 
Combining the form factors (fifth row) with the potential harmonics (third row) yields loop phases $\Phi_q^{\circlearrowleft}$, which determine the general topological phase diagram in the last row. The same colors are used across the entire row for identical values of $C_n$. 
%
}
\label{tab_fullsummary}
\end{table*}

We then include interaction-induced (Hartree-Fock) corrections to the superlattice potential. 
In this context, the remarkable simplification offered by our perturbative symmetry-indicator framework is that the self-consistent Hartree-Fock theory is reduced to a small number of equations -- one for each high-symmetry point --
involving the symmetry eigenvalues of the miniband eigenvectors.
To solve these equations, it is sufficient to loop over the finite different values that they can take, and, for each set of such eigenvalues, perform a simple self-consistency check involving a small number of the material's form factors.
This provides an efficient topological criterion that not only circumvents the computation of the superlattice band structure, 
but also eliminates the need for an effective model to describe the band structure and a full-fledged Hartree-Fock equations for the effective model.

Our approach enables a streamlined and efficient characterization of the topological properties of complex superlattice heterostructures. 
Our work applies to superlattices that are externally imposed by gate nano-patterning~\cite{forsythe2018band,jessen2019lithographic,li2021anisotropic,cano2021moire,wang2021moire,barcons2022engineering,guerci2022designer,sun2023signature,ghorashi2023topological,ghorashi2023multilayer,de2023two,zeng2024gate} or a moir\'e potential arising from proximity coupling to another layer or a substrate with a nearly matched lattice constant and no low energy degrees of freedom~\cite{song2015topological,spanton2018observation,wu2018hubbard,zhang2019nearly,chittari2019gate,chen2020tunable,crepel2023chiral,lu2023fractional,yang2024topological}, as well as to ``anomalous'' superlattices spontaneously generated by electronic interactions~\cite{lu2024fractional,xie2024even,dong2023anomalous,dong2023theory,zhou2023fractional,guo2023theory,song2024intertwined,zeng2024sublattice,dong2024stability}.
While similar ideas have been used before on a case-by-case basis, \textit{e.g.} Refs.~\cite{herzog2023moir,su2022massive,suri2023superlattice,soejima2024anomalous,sarkar2023symmetry}, the present work gives a systematic list of inputs to determine the Chern number analytically (see Tabs.~\ref{tab_fullsummary} and~\ref{tab_summaryhf}). 
These results can be efficiently applied to new moir\'e systems as they arise. 
We illustrate the predictive power of our method by analyzing biased rhombohedral graphene multi-layers in the presence of a nearly-aligned hBN substrate, for which we nail down a remarkable coincidence of length scales responsible for the emergence of the quantum anomalous Hall behaviors observed in penta- and hexa-layer stacks~\cite{lu2024fractional,xie2024even}.

\section{Summary of results}

For clarity of presentation, let us briefly summarize the main results obtained in this manuscript and indicate where they are derived. 
Our starting point is a two-dimensional material or heterostructure, which, upon weak doping, features isolated Fermi pockets centered around some high-symmetry points $\gamma_v$ of its Brillouin zone, with $v$ a valley index labeling the different pockets.  
Our goal is to predict the topological invariant of the mini-bands emerging around $\gamma_v$ when a superlattice potential $V(r)$ with discrete rotation symmetry is applied to the sample, and/or self-consistently generated by interactions. 
As previously mentioned, this is achieved by combining degenerate perturbation theory with the method of symetry indicators. 
We treat the valley $v$ as a good quantum number, which implicitly requires the superlattice to vary slowly on the atomic scale to avoid scattering between the different pockets, and also assume that a single electronic band is relevant near each $\gamma_v$.

\begin{table*}
\centering
\begin{tblr}{colspec = {X[l,m,17cm]}}
\hline 
\texttt{1:} $\;$ Compute a finite number of form factors $\Lambda_{q_2, q_1} = \braket{\chi(q_2)}{\chi(q_1)}$ (12 at most). \\
\texttt{2:} $\;$ For all possible $C_n$ between $0$ and $(n-1)$: \\
\texttt{3:} $\;\qquad$ Evaluate the $\Gamma_{q_0}^{(n)}  [C_n, v, \Lambda]$ (given explicitly in App.~\ref{app_getthegamma}) for all high-symmetry points $q_0$ depicted in Tab.~\ref{tab_fullsummary}. \\
\texttt{4:} $\;\qquad$ If $\left| \arg [ - \Gamma_{q_0}^{(n)} ] \right| < \pi/p(q_0)$ for all $q_0$ (Eq.~\ref{eq_selfconsistentsolutionstronginteraction}): \\
\texttt{5:} $\;\qquad\qquad$ $C_n$ is compatible with a self-consistent Hartree-Fock solution within degenerate perturbation theory. \\
\hline
\end{tblr}
\caption{Efficient determination of all Chern numbers $C_n = C \, ({\rm mod} \, n)$ compatible with a self-consistent Hartree-Fock solution on a reduced Brillouin zone (rBZ) set by an interaction-generated superlattice potential using degenerate perturbation theory. For convenience, we repeat here the definitions given in the main text: $n$ encodes the discrete rotational symmetry $\mathcal{C}_n$ of the problem in the presence of the superlattice, $q_{0,1,2}$ run over the high-symmetry points of the rBZ and their rotated images excluding the rBZ center $\gamma$, and $p(q)$ denotes the largest divisor of $n$ such that $q$ is left invariant by a $2\pi/p$ rotation around $\gamma$ up to reciprocal lattice vectors of the superlattice. The evaluation of each $\Gamma_{q_0}^{(n)}$ requires a handful of elementary operations that only involve the Chern number $C_n$, the form factors $\Lambda_{q_2, q_1}$, and the Fourier transform of the two-body interaction potential $v(q)$ (see Eq.~\ref{eq_definitionsofGammas}).
}
\label{tab_summaryhf}
\end{table*}

We start in Sec.~\ref{sec_noninteracting} with the case of an externally applied superlattice potential that identically couples to all internal degrees of the material or heterostructure under consideration. 
The results of our perturbative analysis in this situation are summarized in Tab.~\ref{tab_fullsummary}, which we quickly describe. 
First, the minimal set of parameters necessary to solve the degenerate-perturbation theory at the high-symmetry points of the reduced Brillouin zone (rBZ) set by the superlattice is identified for all possible rotation symmetry groups $\mathcal{C}_n$ of the system around each valley $\gamma$ (first two rows of Tab.~\ref{tab_fullsummary}). 
These parameters come in two varieties: harmonics of the superlattice potential that resonantly couple states whose degeneracy is imposed by $\mathcal{C}_n$-symmetry, and material-specific form factors between these degenerate states (third and fifth row in Tab.~\ref{tab_fullsummary}, respectively).  
The parameters are combined into gauge invariant quantities, coined loop-phases for reasons detailed in Sec.~\ref{ssec_onehighsym_nonint}, from which the Chern number modulo $n$, $C_n$, of the lowest band can be derived analytically using the method of symmetry indicators~\cite{fang2012bulk} (summarized in Sec.~\ref{ssec_method_nonint}). 
The result of this calculation for each symmetry group, described in Sec.~\ref{ssec_total_nonint}, is presented in the last row of Tab.~\ref{tab_fullsummary}. 
In the limit where the period of the superlattice is long, we can replace the contribution of form factors in the loop phases by a function of $\phi_{\rm B}$, the Berry flux of the original material enclosed by the rBZ. 
This simplifies the analytical formula for $C_n$, derived in Sec.~\ref{ssec_total_nonint} and given in the fourth row of Tab.~\ref{tab_fullsummary}.

These results can be straightforwardly extended to potentials which couple nontrivially to the microscopic degrees of freedom according to a spatially uniform matrix $P$ with the mere replacement $M_q = \braOket{q}{U_{p(q)}}{q} \to \braOket{q}{P U_{p(q)}}{q}$ in the fifth row of Tab.~\ref{tab_fullsummary}, with $p(q)$ the largest divisor of $n$ such that $q$ is left invariant by a $2\pi/p$ rotation around $\gamma$ up to reciprocal lattice vectors of the superlattice, $\ket{q}$ the material's Bloch vectors at momentum $\gamma+q$, and $U_{p}$ the representation of the $(2\pi/p)$-rotation with the same dimension as $\ket{q}$. 
Naturally, the explicit form of $P$ depends on the physical situation considered. 
For concreteness, we demonstrate how this construction works in Sec.~\ref{ssec_pentalayernoint} for the case of a graphene multilayer stack in which the bottom layer is in contact with a nearly-aligned hBN substrate. 
In that case, $P$ represents the projector onto the atoms closest to the substrate. 
Similarly, we can include the perturbative effect of interlayer tunneling in moir\'e homobilayers using a modified version of the $P$ projector, as described in App.~\ref{app_homobilayers}.

In Sec.~\ref{sec_HartreeFock}, we consider Hartree-Fock corrections to the superlattice potential generated by interactions. 
Under a so-called ``resonant'' approximation explained in Sec.~\ref{ssec_resonantapprox}, these corrections can be computed from the sole knowledge of the form factors between all high-symmetry points of the rBZ and the symmetry eigenvalues at these high-symmetry points. 
This allows us to derive self-consistent equations for the symmetry eigenvalues that only depend on a few material-specific form factors and the harmonics of the externally applied superlattice potential. 
Within our formalism, the self-consistent nature of these equation only adds a minimal computational cost to the determination of the Chern number. 
Indeed, the symmetry eigenvalues can only take a finite number of values, over which we can loop to compute the Hartree-Fock corrections and check whether they give rise to self-consistent solutions for the symmetry eigenvalues at all high-symmetry points. 
When interactions effects dominate the physics of the system under consideration, \textit{i.e.} if the amplitude of the superlattice is negligible, this loop over symmetry eigenvalues simplifies to a loop over the $n$ possible Chern numbers (modulo $n$), as shown in Sec.~\ref{ssec_stronginteractionHF} and sketched as an algorithm in Tab.~\ref{tab_summaryhf}. 
This algorithm only necessitates the evaluation of auxiliary parameters (the $\Gamma_{q_0}^{(n)}$ defined in Eq.~\ref{eq_definitionsofGammas}) that depend on a finite set of material form factors and on the two-body interaction potential.

Finally, in Sec.~\ref{sec_multilayergraphene}, we apply our formalism to biased rhombohedral graphene multi-layers stacked above a nearly-aligned hBN substrate, in which quantum anomalous Hall behaviors have recently been experimentally observed~\cite{lu2024fractional,xie2024even}. 
Not only does our perturbative topological criterion correctly capture the phases observed experimentally, but it also offers a unique analytical insight into the remarkable coincidence responsible for the observed quantum anomalous Hall behaviors. 
More precisely, a careful inspection of the algorithm in Tab.~\ref{tab_summaryhf} shows the transition to a quantum anomalous Hall phase with Chern number $C_3 = -1$ (mod 3) when the phase of the Fock term reaches a value $> \pi/3$. 
This phase only depends on the form factors of the graphene stack, and is directly proportional to the flux of Berry curvature threading the rBZ. 
Computing this flux, we find that the transition to a topological insulating state occurs when the superlattice period $a_{\rm super}^* \simeq  \frac{v_F K a_0}{0.6 t_1} = \SI{11}{\nano\meter}$, where $v_F$, $K$ and $a_0$ respectively denote the Fermi velocity, Brillouin zone corner and lattice constant of the graphene multilayer, while $t_1$ represents the nearest-neighbor interlayer tunneling amplitude within the stack. 
The exceptional coincidence explaining the experimental observation of quantum anomalous Hall behaviors in Refs.~\cite{lu2024fractional,xie2024even} is that, for a graphene/hBN twist of about $1^\circ$, the superlattice potential imposed by the hBN substrate lies very close to this special value $a_{\rm super}^*$.

\section{Non-interacting theory} \label{sec_noninteracting}

In this section, we introduce our methodology for determining the Chern number of the lowest-energy band in the rBZ resulting from an externally applied superlattice potential $V(r)$, discarding the effects of interactions that will be included in Sec.~\ref{sec_HartreeFock}.

\subsection{Method} \label{ssec_method_nonint}

Let us focus on a single valley $\gamma$ and describe the method of symmetry indicators for determining the topological content of the superlattice-induced minibands. 
To do this, we first introduce some notations. 
We respectively call $\varepsilon_{q}$ and $\ket{\chi (q)}$ the energy and the periodic part of the Bloch eigenvector at momentum $k = \gamma + q$ of the band giving rise to the Fermi pocket at $\gamma$. 
We decompose the superlattice potential into Fourier components 
\begin{equation} \label{eq_superlattice}
V(r) = - \sum_{g} V_g e^{i g\cdot r} ,  
\end{equation}
where $V_{-g} = V_g^*$ is required for the potential to be real, while the overall negative sign is conventional and chosen to recall that electrons have negative electric charge. 
The quasi-momentum $k$ in the original band spans the full Brillouin zone (BZ) of the original material, which is folded due to the superlattice potential into a reduced Brillouin zone (rBZ) defined by the $g$-vectors.
We denote by $\mathcal{C}_n$ the largest subgroup of the point symmetry group at $\gamma$ under which the potential $V(r)$ is invariant. 
Under $R_n$, the rotation by $2\pi/n$, it follows that $\varepsilon_{R_n q} = \varepsilon_{q}$ and $V_{R_n g} = V_{g}$. 
We choose a gauge in which the action of $R_n$ on the Bloch wavevectors is given by a representation $U_n$ which satisfies 
\begin{equation}
U_n \ket{\chi(q)} = \lambda_n(\gamma) \ket{\chi (R_n q)} , \quad \lambda_n(\gamma) = e^{\frac{2i\pi}{n} \nu_n(\gamma)} , 
\end{equation}
for all $q$ in the rBZ, where $\lambda_n(\gamma)$ is the $R_n$ eigenvalue of $\ket{\chi(0)}$. We take $\nu_n(\gamma)$ to be an integer, which is valid in the absence of spin-orbit coupling.

With these definitions, the Hamiltonian at momentum $k = \gamma + q$ projected in the band of interest takes the form 
\begin{equation} \label{eq_superlatticehamiltonian}
H (q) = \sum_g \varepsilon_{q+g} c_{q+g}^\dagger c_{q+g} - V_g \Lambda_{q+g,q} c_{q+g}^\dagger c_q ,
\end{equation}
where $c_q$ is the fermionic annihilation operator corresponding to $\ket{\chi (q)}$, the form factor $\Lambda_{q',q} = \braket{\chi(q')}{\chi(q)}$ arises from projecting onto a single band, and $q$ runs over the BZ.  
The gauge chosen above makes clear that these form factors, and hence $H(q)$, are $\mathcal{C}_n$-symmetric.

To determine whether the lowest energy miniband of Eq.~\ref{eq_superlatticehamiltonian} has non-trivial topology, we rely on the symmetry indicators developed in Ref.~\cite{fang2012bulk}. 
There, it is proved that the Chern number $C$ of an isolated band with $R_n$ symmetry can be determined modulo $n$ by the product of $\lambda_p(q_p)$, the $R_p$-eigenvalues of the band's Bloch eigenvectors at momenta $q_p$ of the rBZ 
invariant under $R_p$,
where $p$ runs over all divisors of $n$. 
Because $n=2,3,4,6$ can only take four values in two dimensions, this result can be written explicitly as
\begin{equation} \label{eq_ChernFromSym}
\begin{cases}
e^{\frac{2i \pi}{2} C_{2}} = \lambda_2(\gamma) \lambda_2(m) \lambda_2(x) \lambda_2(y) & \text{for}\, n=2 \\
e^{\frac{2i \pi}{3} C_{3}} = \lambda_3(\gamma) \lambda_3(\kappa) \lambda_3(\kappa') & \text{for}\, n=3 \\
e^{\frac{2i \pi}{4} C_{4}} = \lambda_4(\gamma) \lambda_4(m) \lambda_2(x) & \text{for}\, n=4 \\
e^{\frac{2i \pi}{6} C_{6}} = \lambda_6(\gamma) \lambda_3(\kappa) \lambda_2(m)  & \text{for}\, n=6 
\end{cases} ,
\end{equation}
with $C_{n} = C  \, ({\rm mod} \, n)$, and where the definition of the high-symmetry points is given graphically in the second line of Tab.~\ref{tab_fullsummary} for all $n$. 
We have discarded the spin contribution, which does not play any role in absence of spin-orbit coupling.

In the rest of this section, we determine whether the lowest superlattice induced miniband from the $\gamma$-valley is topological using the symmetry indicator formulas (Eq.~\ref{eq_ChernFromSym}) by analytical solution of the band-projected superlattice Hamiltonian (Eq.~\ref{eq_superlatticehamiltonian}) at the high-symmetry points depicted in the second line of Tab.~\ref{tab_fullsummary}, which is possible when $V(r)$ is treated as a perturbation. 
The topological invariants obtained by this degenerate perturbation theory remain valid for non-perturbative amplitudes of the superlattice potential as long as the gaps opened for small $V(r)$ do not close when the amplitude of the superlattice is increased to its full value.

\subsection{A single high-symmetry point} \label{ssec_onehighsym_nonint}

We now compute the eigenvalues appearing in Eq.~\ref{eq_ChernFromSym} for the lowest energy band formed by the superlattice at all the high-symmetry points depicted in the second line of Tab.~\ref{tab_fullsummary}. 
As explained above, we compute these eigenvalues assuming a weak superlattice potential, which enables an analytical solution of Eq.~\ref{eq_superlatticehamiltonian} using degenerate perturbation theory. 
The obtained eigenvalues remain unchanged for non-perturbative amplitudes of the superlattice potential as long as the gaps opened for small $V(r)$ do not close as $V(r)$ is increased to its full value. 
The results obtained here will be used in Sec.~\ref{ssec_total_nonint} to predict which combinations of superlattice geometry and form factors yield topological minibands.

Let us consider a high-symmetry point $q_0 \neq \gamma$ in the rBZ with point group symmetry $\mathcal{C}_p$, where $p$ necessarily divides $n$. 
We discard $\gamma$ from the present discussion, since its $\mathcal{C}_n$-eigenvalue $\lambda_n(\gamma)$ is a property of the material under consideration that does not change under a weak superlattice potential. 
Discarding $\gamma$ restricts $p \in \{2,3,4\}$ to only three distinct values (see Eq.~\ref{eq_ChernFromSym}). 
The diagonalization of $H(q_0)$ for weak superlattice potential can be obtained by projecting the Hamiltonian into the subspace of degenerate eigenvectors with momenta $q$ related by $\mathcal{C}_p$ symmetry to $q_0$, \textit{i.e.} in the basis $[\ket{\chi ( q )}, q\in \mathcal{B}(q_0)] $ with  
\begin{equation} \label{eq_projectionbasis}
\mathcal{B}(q_0) = \{ q_k = R_n^{k n/p} q_0 = R_p^k q_0 \, | \, k=0, \cdots , p-1 \} . 
\end{equation}
The Hamiltonian projected onto the degenerate subspace $\mathcal{B}(q_0)$ takes the circulant form 
\begin{equation} \label{eq_circulantform}
P H(q_0) P = \begin{bmatrix}
a_0 & a_{p-1} & \cdots & a_2 & a_1 \\
a_1 & a_0 & a_{p-1} &  & a_2 \\
\vdots & a_1 & a_0 & \ddots & \vdots \\
a_{p-2} & & \ddots & \ddots & a_{p-1} \\ 
a_{p-1} & a_{p-2} & \cdots & a_1 & a_0
\end{bmatrix} ,
\end{equation}
with $P$ the projector onto the basis of Eq.~\ref{eq_projectionbasis}, and where
\begin{equation}
a_0 = \varepsilon_{q_0} , \quad a_{k>0}  = - V_{q_k - q_0} \Lambda_{q_k , q_0} = a_{p-k}^* . 
\end{equation}
The eigenvalues and eigenvectors of $P H(q_0) P$ are obtained by discrete Fourier transform, and read
\begin{equation} \label{eq_diagbyfouriertransform}
E_\nu = \sum_{k=0}^{p-1} \omega_p^{\nu k} a_k, \quad \ket{\phi_\nu} = \sum_{k=0}^{p-1} \frac{ \omega_p^{-\nu k} \ket{\chi(q_k)} }{\sqrt{p}}  , \quad \omega_p = e^{\frac{2i\pi}{p}} ,
\end{equation}
from which it is clear that $\ket{\phi_\nu}$ has $\mathcal{C}_p$-eigenvalue 
\begin{equation} \label{eq_Cpeigs}
\braOket{\phi_\nu}{U_p}{\phi_\nu} = [\lambda_n(\gamma)]^{n/p} \omega_p^{\nu} , \quad  U_p = (U_n)^{\frac{n}{p}} .
\end{equation}

The $\mathcal{C}_p$-eigenvalues in Eq.~\ref{eq_Cpeigs} belong to the $p$ lowest-energy minibands at $q_0$ in the rBZ. To apply the symmetry indicator formula to the \textit{lowest} miniband, we need the $\mathcal{C}_p$-eigenvalue of that specific band. Thus, we now determine the smallest $E_\nu$. 
We first rewrite the energies in Eq.~\ref{eq_diagbyfouriertransform} using the relation $a_{k} = a_{p-k}^*$ to find
\begin{equation}
E_\nu = \varepsilon_{q_0} + \begin{cases} \Re [\omega_2^\nu a_1] & p=2 \\  \Re [2 \omega_3^\nu a_1] & p=3 \\  \Re [2 \omega_4^\nu a_1 + \omega_2^\nu a_2] & p=4  \end{cases} . 
\label{eq_energies}
\end{equation}
Under the assumption that the largest harmonics of the potential dominate over the second one when $p=4$, \textit{i.e.} when $a_2=0$, the energy $E_\nu$ is minimized for all $p$ when the argument of $\omega_p^\nu a_1$ is closest to $\pi$ modulo $2\pi$. 
Using $\arg [\omega_p^\nu a_1] = \pi + \arg(V_{q_1 - q_0} \Lambda_{q_1,q_0}) + 2\nu \pi/p$, the energy is minimized when the integer $\nu$ is closest to $[ - p \arg(V_{q_1 - q_0} \Lambda_{q_1,q_0}) / 2\pi ]$ modulo $p$. This is realized for 
\begin{equation} \label{eq_nupforq0}
\nu_p (q_0) = \left\lfloor \frac{\pi - \Phi_{q_0}^{\circlearrowleft}}{2\pi}  \right\rfloor , \quad \Phi_{q_0}^{\circlearrowleft} = p \arg ( V_{g_{q_0}} \Lambda_{q_1,q_0} ) ,
\end{equation}
where $\lfloor x \rfloor$ denotes the largest integer smaller than $x$, and $g_{q_0} = q_1-q_0$. 
Let us highlight that the phase $\Phi_{q_0}^{\circlearrowleft}$ can also be written as
\begin{equation} \label{eq_Phi2} 
\Phi_{q_0}^{\circlearrowleft}  = p \arg \left[ V_{g_{q_0}}  \frac{ \braOket{\chi(q_0)}{U_p^\dagger}{\chi(q_0)} }{ \braOket{\chi(\gamma)}{U_p^\dagger}{\chi(\gamma)} }\right] ,
\end{equation}
which only involves expectation values of Bloch eigenvectors that are local in momentum space and can therefore be efficiently computed from results of a density functional theory or tight-binding calculation of the original material. This is the form of the loop phases given in Tab.~\ref{tab_fullsummary}, where we have used a different symbol
\begin{equation}
M_{q_0} = \braOket{\chi(q_0)}{U_p}{\chi(q_0)}  ,
\end{equation}
for the form factors to make clear that the formula provided is independent of the gauge chosen for the Bloch vectors. Only in the $C_n$-symmetric gauge chosen at the beginning of the section do we have $M_{q_0} = \Lambda_{q_0,q_1}$.

\begin{table}
\centering
\begin{tblr}{colspec = {X[c,m,1.25cm]||X[c,m,1.25cm]|X[c,m,0.8cm]|X[c,m,2.5cm]}}
$p$ & $q_0$ & $r_p(q_0)$ & $(q_k)_{k=0,\cdots, p}$ loop  \\ \hline\hline
2, 3, 4, 6 & $\gamma$ & 0 & \loopgamma \\  \hline
2 & $x$, $y$, $m$ & 0 & \looppequaltwo \\ \hline
3 & $\kappa$, $\kappa'$ & 1/2 & \looppequalthree \\ \hline
4 & $m$ & 1 & \looppequalfour 
\end{tblr}
\caption{
Geometric factor $r_p(q_0)$ measuring the fraction of the rBZ spanned by the $(q_k)_{k=0,\cdots, p}$ loop for all high-symmetry points $q_0$ appearing in Eq.~\ref{eq_ChernFromSym}, where the point group symmetry at $q_0$ is $\mathcal{C}_p$. The last column shows a representative loop for each case considered, where the area enclosed by the loop is colored in red while the rBZ is shown with a solid black line. In the first two cases, this area should be understood to vanish as the loop contracts to a point or to a line.
}
\label{tab_geometricfactor}
\end{table}

The phases $\Phi_{q_0}^{\circlearrowleft}$ have a simple physical interpretation: they correspond to the total phase acquired along the discrete loop $\{q_k = R_p^k q_0 \}_{k=0,\cdots , p}$ connecting degenerate high-symmetry points, including contributions from both the potential harmonics and the Berry flux enclosed within the loop (explaining the $\circlearrowleft$ superscript). 
To see this, use the $\mathcal{C}_p$ symmetry to rewrite
\begin{equation} \label{eq_PhiAsLoop}
\Phi_{q_0}^{\circlearrowleft} = \arg \left[ \prod_{k=0}^{p-1} V_{q_{k+1} - q_k} \Lambda_{q_{k+1}, q_k} \right] .
\end{equation}
This manifestly gauge-invariant form allows to interpret $\Phi_{q_0}^{\circlearrowleft}$ as the phase acquired along the discrete loop connecting $q_0$ to its $R_p$-rotated images (see Tab.~\ref{tab_geometricfactor} for illustrations). In practice, Eq.~\ref{eq_PhiAsLoop} should also be used when the gauge chosen in a particular calculation is not $\mathcal{C}_p$-symmetric, as assumed when writing Eq.~\ref{eq_eigvalsfull} (recall the discussion of gauge choice above Eq.~\ref{eq_superlatticehamiltonian}).

In the limit where the period of the superlattice is long --- more precisely, when the primitive reciprocal lattice vectors of the rBZ are small compared to the typical scale on which the quantum geometry of the original band varies --- the product of form factors in Eq.~\ref{eq_PhiAsLoop} gives the total Berry flux of the original material enclosed within the loop, and we can write
\begin{equation} \label{eq_berryphaseforoverlap}
\Phi_{q_0}^{\circlearrowleft} \simeq p \arg [ V_{g_{q_0}} ] + r_p (q_0) \phi_{\rm B} ,  \quad \phi_{\rm B} = \Omega_\gamma A_{\rm rBZ} , 
\end{equation}
with $\phi_{\rm B}$ the total Berry flux enclosed by the rBZ, written as the product of $\Omega_\gamma$, the Berry curvature at $\gamma$, with the rBZ area, $A_{\rm rBZ}$. The geometrical factor $r_p(q_0)$ denotes the fraction of the rBZ area enclosed by the $(q_k)_{k=0,\cdots, p-1}$ loop. 
The values of this geometric factor for the different high-symmetry points are tabulated in Tab.~\ref{tab_geometricfactor}. 
When $p$ is even, $R_2$ is a symmetry element, which implies $V_{g} = V_{R_2 g = -g} = V_{g}^*$, \textit{i.e.}, $V_{g}$ is real. Thus, we can substitute
\begin{equation} \label{eq_realpotentialC2}
\arg (V_{g}) = \frac{\pi}{2} [1-\sign (V_{g})] = \pi \Theta(-V_g) 
\end{equation}
in the definition of $\Phi_p^{\circlearrowleft}(q_0)$, where $\sign(x)$ and $\Theta(x)$ respectively denote the sign and Heavyside functions.

Combining Eqs.~\ref{eq_Cpeigs} and~\ref{eq_nupforq0}, we finally obtain
\begin{equation} \label{eq_eigvalsfull}
\lambda_p(q_0) = e^{\frac{2i\pi}{p} [\nu_n(\gamma) + \nu_p (q_0)]} . 
\end{equation}
This is the first main result of this paper. 
It relates the symmetry eigenvalues $\lambda_p(q_0)$ of the lowest superlattice-induced miniband to the gauge-invariant loop phase $\Phi_{q_0}^{\circlearrowleft}$, which includes the effect of the potential and of the form factors of the material. 
These loop phases can be efficiently computed from the sole knowledge of the superlattice geometry, the Bloch states at high-symmetry points, and the matrix representation $U_n$ of the $\frac{2\pi}{n}$-rotation $R_n$ in their Hilbert space. 
In the long-period limit, the dependence on the form factors is entirely carried by $\phi_{\rm B}$,  the total Berry flux of the original material threading the reduced Brillouin zone defined by the superlattice potential. 
Together, Eq.~\ref{eq_ChernFromSym} and Eq.~\ref{eq_eigvalsfull} reduce the problem of computing the Chern number of the lowest superlattice-induced miniband to the calculation of a finite number of matrix elements, as we will carry out in the next section.

\subsection{Topological invariants} \label{ssec_total_nonint}

We now combine the expression for the Chern number in terms of symmetry eigenvalues with our perturbative calculation of the latter to determine conditions under which the lowest superlattice-induced miniband acquires a Chern number. Using Eq.~\ref{eq_eigvalsfull}, the Chern number modulo $n$ in Eq.~\ref{eq_ChernFromSym} formally becomes
\begin{equation} \label{eq_formalcombination} \begin{split}
C_n & = \nu_n (\gamma) + \sum_{q_0\neq \gamma} \frac{n}{p(q_0)} \left[ \nu_n (\gamma) + \nu_p (q_0) \right] \\ & = \sum_{q_0\neq \gamma} \frac{n}{p(q_0)}\nu_p (q_0) 
 , 
\end{split} \end{equation}
where $q_0$ runs over the high-symmetry points of the Brillouin zone excluding $\gamma$, $\nu_p(q_0)$ is given in Eq.~\ref{eq_nupforq0}, and $p(q_0)$ is the largest divisor of $n$ for which $R_n^{n/p} q_0 = q_0$. 
To obtain the second line in Eq.~\ref{eq_formalcombination}, we have removed a multiple of $n \nu_n (\gamma)$ from the first line, since $C_n$ is defined modulo $n$.

In Appendix~\ref{app_combineeigvals}, we treat the different values of $n=2,3,4,6$ separately, identify the independent variables that determine $C_{n}$ in each case, and simplify the formula for $C_{n}$ in the long-period limit (discussed above Eq.~\ref{eq_berryphaseforoverlap}). 
The results populate Tab.~\ref{tab_fullsummary}, which provides a concise summary of our non-interacting results.

As an illustrative example applying our results, we now derive the recent observation, made in several recent works~\cite{ghorashi2023topological,scheer2023kagome,tan2024designing,zeng2024gate,zeng2024sublattice}, that superlattices with honeycomb geometry have a stronger propensity to form topological minibands than triangular ones. 
To see this, let us focus on the $n=3$ column of Tab.~\ref{tab_fullsummary} and rely on the long-wavelength formula for the Chern number $C_3 = C$ (mod $3$). 
We first consider a triangular superlattice ($\theta=0$), which, in absence of Berry curvature at $\gamma$ in the original material ($\phi_{\rm B}=0$), yields loop phases equal to zero for both $\kappa$ and $\kappa'$. 
This corresponds to the origin of the $n=3$ topological phase diagram in Tab.~\ref{tab_fullsummary} (last row), where $C_3 = 0$. 
The addition of Berry curvature at $\gamma$ moves both loop phases by the same amount, \textit{i.e.} along the first diagonal of the topological phase diagram, such that the system remains in the $C_3 = 0$ phase even for non-zero values of $\phi_{\rm B}$. 
In fact, the long-wavelength analytic formula shows that reaching a state with non-zero Chern number requires the Berry flux through the rBZ to reach $| \phi_{\rm B} | >2\pi$, which is highly unlikely for small rBZ. 
On the other hand, we observe that honeycomb superlattices ($\theta=\pi$) favor topological states for any infinitesimal amount of enclosed Berry curvature $\phi_{\rm B}$. 
This can be seen in the $n=3$ topological phase diagram in Tab.~\ref{tab_fullsummary}, where the honeycomb case with $\phi_{\rm B}=0$ corresponds to the $(\pi, \pi)$ point, which is surrounded along the first diagonal by phases with $C_3 \neq 0$. 
Adding Berry curvature necessarily drives the system into one of those topological phases. 
This can also be seen by plugging $\theta =\pi$ in the long-wavelength formula of Tab.~\ref{tab_fullsummary}, yielding $C_3 = 1 + 2 \lfloor \frac{-\phi_{\rm B}}{4\pi} \rfloor$ that is either equal to $1$ ($\phi_{\rm B}<0$) or to $-1$  ($\phi_{\rm B}>0$) for $|\phi_{\rm B}|<4\pi$.

\section{Effects of interactions} \label{sec_HartreeFock}

The applicability of the results derived in Sec.~\ref{sec_noninteracting} relies on two main assumptions. 
The first, explicitly stated in Sec.~\ref{ssec_method_nonint},
is that the gaps opened by the superlattice potential at its full strength do not close as the potential strength is turned down to a perturbatively small value.
The second, implicit in the single-particle formalism of the last section, is that interactions do not contribute to the gap opening at the high-symmetry points. 
In this section, we relax this second hypothesis and demonstrate that the Chern number of the lowest miniband can still be determined from a small number of form factors of the original band structure in the presence of interactions when the interactions are treated perturbatively. 
As in the previous sections, the results obtained in this perturbative limit remain correct if the gaps at high-symmetry points of the rBZ do not close as the superlattice amplitude and interaction strength are tuned to their full values. 

\subsection{Self-consistent Hartree-Fock equations} \label{ssec_resonantapprox}

\subsubsection{Hartree-Fock potentials}

We consider a positive two-body interaction potential $v(q)>0$ with continuous rotation symmetry, and project it onto the relevant band around $\gamma$ to obtain, within the Hartree-Fock approximation, an updated form of the Hamiltonian in Eq.~\ref{eq_superlatticehamiltonian}
\begin{equation} \label{eq_superlatticehamiltonianHF}
H(q) = \sum_g \varepsilon_{q+g} c_{q+g}^\dagger c_{q+g} +[ V_{q,g}^{\rm HF} - V_g ] \Lambda_{q+g,q} c_{q+g}^\dagger c_q , 
\end{equation}
where the interaction-induced part of the effective potential $V^{\rm HF}$ arises from non-zero expectation values $\rho_{k,g} = \Lambda_{k+g,k} \langle c_{k+g}^\dagger c_k \rangle$ in the ground state, and reads
\begin{subequations} \label{eq_HartreeFock} \begin{align}
V_{q,g}^{\rm HF} & = V_{g}^{\rm H} - V_{q,g}^{\rm F} , \quad V_{g}^{\rm H} =  \int \frac{{\rm d}^2 k}{A_{\rm BZ}}  v(g) \rho_{k,g}^* , \\
V_{q,g}^{\rm F} & =  \int' \frac{{\rm d}^2 k}{A_{\rm BZ}} \frac{\Lambda_{k,q} \Lambda_{k+g,q+g}^*}{\Lambda_{q+g,q} \Lambda_{k+g,k}^*} v(k-q) \rho_{k,g}^* ,
\end{align} \end{subequations}
with $A_{\rm BZ}$ the area of the original material's Brillouin zone. Here, $V_{g}^{\rm H}$ and $V_{k,g}^{\rm F}$ respectively correspond to the Hartree and Fock potentials. 
We assume that they do not spontaneously break the $\mathcal{C}_n$ rotation symmetry of the system, \textit{i.e.} $V_{R_n g}^{\rm H} = V_{g}^{\rm H}$ and $V_{R_n q, R_n g}^{\rm F} = V_{q,g}^{\rm F}$. 
Notice that the ratio of form factors in the Fock term is gauge invariant; its phase corresponds to the Berry flux accumulated around the momentum-space loop $(q \to k \to k+g \to q+g \to q)$. 
Finally, we do not consider the zero-momentum exchange terms of the interactions as they only contribute a global constant energy shift proportional to the square of the total density. This excludes the $k=q$ part of the Fock integral, which we have made explicit using a primed integral notation.

Our goal is to find exact solutions of $H(q)$ at all high-symmetry points $q_0$ of the rBZ in the perturbative regime, for which the Hamiltonian can be projected onto the set of states $\{ \ket{\chi(k)} \, |\, k\in\mathcal{B}(q_0) \}$ whose exact degeneracy is enforced by $\mathcal{C}_n$-symmetry (as explained in Sec.~\ref{sec_noninteracting}). 
Similar to the non-interacting case, the projected Hamiltonian $PH(q)P$ takes the circulant form of Eq.~\ref{eq_circulantform} with the modified parameters
\begin{equation}
a_0 = \varepsilon_{q_0} , \quad a_k = \left[ V_{q_0, q_k - q_0}^{\rm HF} - V_{q_k - q_0} \right] \Lambda_{q_k , q_0} , 
\end{equation}
and the derivation of the symmetry eigenvalues in Sec.~\ref{ssec_onehighsym_nonint} follows, yielding
\begin{equation} \label{eq_nupHF} \begin{split}
\nu_p (q_0) & = \left\lfloor \frac{\pi - \Phi_{q_0}^{\circlearrowleft}}{2\pi}  \right\rfloor , \\ \Phi_{q_0}^{\circlearrowleft} & = p \arg [ (V_{q_1-q_0} - V_{q_0, q_1-q_0}^{\rm HF}) \Lambda_{q_1,q_0} ] .
\end{split} \end{equation}
However, in contrast to the non-interacting case, this equation is not sufficient to determine the symmetry eigenvalue $\nu_p (q_0)$, because the Hartree-Fock potential $V^{\rm HF}$ formally depends on the value of $\rho_{k,g}$ for all $k$ in the original BZ. 
To make analytical progress, we now resort to an approximation, justified in the perturbative regime considered throughout this manuscript, to simplify the functional of the Hartree-Fock potential. 

\subsubsection{Resonant approximation}

What justifies the restriction to high-symmetry points 
in Sec.~\ref{sec_noninteracting} is that, for the weak perturbation considered, the original band-structure of the material remains unchanged except in small neighborhoods around the rBZ high-symmetry points. In these small patches, there exist superlattice harmonics with momentum $g$ that resonantly couple states of the original band with the same energy (their degeneracy is enforced by the $\mathcal{C}_n$ rotation symmetry). 
In other words, the state $\ket{\chi(k)}$ from the original band structure only substantially hybridizes with $\ket{\chi(k+g)}$ if $k$ lies close enough to a high-symmetry point $q_0$ for which $q_0+g \in \mathcal{B}(q_0)$.
This leads to non-zero values of $\rho_{k,g} = \Lambda_{k+g,k} \langle c_{k+g}^\dagger c_k \rangle$ only in those regions where the superlattice potential is resonant. 
Our perturbative treatment only holds when the area $A_{\rm res}$ of resonant regions remains small compared to the total Brillouin zone area, as measured by the coefficient $\eta = A_{\rm res} / A_{\rm BZ} \ll 1$.
Because of this, we can assume $\rho_{k,g}$ and the form factors appearing in the Fock integral are independent of momentum within each of the small patches of size $A_{\rm res}$. 
Altogether, this allows to replace the integral in the definition of $V_{q,g}^{\rm HF}$ by the sum
\begin{equation} \label{eq_reductionintegralstosums}
\int \frac{{\rm d}^2 k}{A_{\rm BZ}} \simeq \eta \sum_{(k, k+g) \in \mathcal{B}(q_0)} ,
\end{equation}
an approximation that we refer to as the ``resonant approximation.''  
We point out that this resonant approximation becomes equivalent to the few-patch model introduced in Ref.~\cite{dong2023anomalous} when the resonance region $A_{\rm res}$ is assumed to be as large as the rBZ.

\subsubsection{Self-consistency condition}

Under the resonant approximation described above, the only order parameters $\rho_{k,g}$ appearing in the Hartree-Fock integrals involve momenta $(k,k+g) \in \mathcal{B}(q_0)$, \textit{i.e.,} rotated images of a high-symmetry point $q_0$. Thus, there exist integers $\alpha, \beta$ such that $k = R_p^\alpha q_0$ and $k+g = R_p^\beta q_0$. Using the analytic form of the lowest miniband eigenvector $\ket{\phi_{\nu_p(q_0)}}$ at $q_0$ derived in Eq.~\ref{eq_diagbyfouriertransform}, we find
\begin{equation} \label{eq_selfconsistentrho} \begin{split}
\rho_{k,g} & = \Lambda_{k+g, k} \braOket{\phi_{\nu_p(q_0)}}{ c_{k+g}^\dagger c_{k} }{\phi_{\nu_p(q_0)}} \\ & = \frac{1}{p} \Lambda_{k+g, k} \, \omega_p^{(\beta-\alpha)\nu_p(q_0)} .
\end{split} \end{equation}
This equation gives the order parameters appearing in the Hartree-Fock potential as a function of the symmetry eigenvalue and a finite set of form factors of the original band. 
In App.~\ref{app_getthegamma} we combine Eqs.~\ref{eq_reductionintegralstosums} and \ref{eq_selfconsistentrho} to rewrite the Hartree-Fock potential at all high-symmetry points $q$ as 
\begin{equation} \label{eq_simplifiedHFintermsofGamma}
V_{q,g}^{\rm HF} = \frac{\eta v(g)}{p} \omega_p^{-\nu_p(q)} \Lambda_{q,q+g}  \Gamma_{q}^{(n)} (C_n , v, \Lambda ) ,
\end{equation}
where the auxiliary coefficients $\Gamma_{q}^{(n)}(C_n, v, \Lambda )$ only depend on the Chern number modulo $n$ of the lowest miniband $C_n$, the two-body interaction potential $v(q)$, and a finite number of form factors of the original band $\Lambda_{q_1,q_0}$ with both $q_0$ and $q_1$ high-symmetry points of the rBZ. The analytical expressions of the $\Gamma_{q}^{(n)}$ are given for all $n=2,3,4,6$ and all high-symmetry points $q$ in App.~\ref{app_getthegamma}.

Eqs.~\ref{eq_nupHF} and~\ref{eq_simplifiedHFintermsofGamma} form the second main result of this paper: the topology of the (fully filled) lowest miniband can be perturbatively determined from a few form factors of the original band structure as the solution to self-consistent equations for the symmetry eigenvalues at high-symmetry points. 
Contrary to the non-interacting case presented in Sec.~\ref{sec_noninteracting} however, these self-consistent equations cannot be solved analytically to provide a comprehensive phase diagram akin to those from Tab.~\ref{tab_fullsummary}. 
Nonetheless, because the symmetry eigenvalues can only take a finite number of values, one can solve the equations algorithmically by looping over all allowed eigenvalues and checking whether the self-consistency condition of Eq.~\ref{eq_nupHF} is satisfied by computing Eq.~\ref{eq_simplifiedHFintermsofGamma} using a finite number of the original material's form factors. 
These steps are described in Tab.~\ref{tab_summaryhf} in the regime where the amplitude of the Hartree-Fock potential is greater than that of the externally applied superlattice, which we now investigate in more detail.

\subsection{Interaction-driven topology} \label{ssec_stronginteractionHF}

A subtle issue introduced by the resonant approximation described in the previous paragraph is the emergence of the phenomenological parameter $\eta$ measuring the size of the resonant area. 
Fortunately, this parameter drops out
in the limit where the strength of the Hartree-Fock potential dominates over the amplitude of the applied superlattice potential. 
In this case, we replace $V_{q,g}^{\rm HF} - V_g \simeq V_{q,g}^{\rm HF}$ in the loop phases, which gives
\begin{equation} \begin{split}
\Phi_{q_0}^{\circlearrowleft} & \simeq p \arg \left[ - V_{q_0, q_1-q_0}^{\rm HF} \Lambda_{q_1,q_0} \right] \\ & = p \arg \left[ -\Gamma_{q_0}^{(n)} \right] - 2\pi \nu_p(q_0)
\end{split} \end{equation}
independent of the phenomenological parameter $\eta$. 
This drastically simplifies the self-consistent equations, which are recast as
\begin{equation} \label{eq_selfconsistentsolutionstronginteraction}
\left\lfloor \frac{\pi - p (q_0) \arg \left[ - \Gamma_{q_0}^{(n)} \right] }{2\pi}  \right\rfloor = 0 , 
\end{equation}
where, as in Sec.~\ref{sec_noninteracting}, $p(q_0)$ denotes the largest divisor of $n$ for which $R_n^{n/p} q_0 = q_0$.
A self-consistent Hartree-Fock solution requires Eq.~\ref{eq_selfconsistentsolutionstronginteraction} to be simultaneously satisfied for all high-symmetry points $q_0$ of the rBZ.

In the limit where the superlattice potential is weak compared to the interaction scale, checking whether a self-consistent Hartree-Fock solution with Chern number $C_n$ exists simply reduces to the computation of $\Gamma_{q_0}^{(n)}$ for all high-symmetry points $q_0$ of the rBZ using a few form factors $\Lambda$ of the original band structure and the functional form of the two-body interaction potential $v(q)$, in order to check that Eq.~\ref{eq_selfconsistentsolutionstronginteraction} is satisfied. 
This constitutes an efficient algorithm to predict which Chern number is favored by interactions solely from the knowledge of the few form factors appearing in Eqs.~\ref{eq_definitionsofGammas}, as summarized in Tab.~\ref{tab_summaryhf} in the introduction. 
When no superlattice potential is present and when the form factors of the material do not yield any self-consistent solutions for the symmetry eigenvalues $\nu_p(q_0)$,
at least on of the high-symmetry points considered remains gapless in the presence of perturbatively weak interactions. This crossing can still be opened for larger values of the interaction amplitude in the strong-coupling limit.

\section{Application to rhombohedral graphene multi-layers} \label{sec_multilayergraphene}

In this section, we apply our formalism to rhombohedral (also referred to as $ABC-$stacked) graphene multi-layers. This choice is motivated by the recent discovery of quantum anomalous Hall effects in electrically biased penta- and hexa-layer graphene subject to a superlattice potential induced on its bottom surface by a near-aligned hexagonal boron nitride (hBN) substrate~\cite{lu2024fractional,xie2024even}. 
We will show that the theory developed in the previous sections captures these experimental observations, providing an elegant and numerically efficient route to the topological phase diagram of the graphene $N$-layer stacks. 
We organize our discussion as follows: Sec.~\ref{ssec_modelpentalyaer} introduces the tight-binding model used in our calculations; Sec.~\ref{ssec_pentalayernoint} applies the non-interacting theory of Sec.~\ref{sec_noninteracting} and highlights that the effect of the substrate-induced superlattice potential decreases exponentially fast as a function of the bias applied across the sample; Sec.~\ref{ssec_pentalayerincludeint} includes the effects of interactions using the methodology given in Sec.~\ref{sec_HartreeFock} to obtain the Hartree-Fock phase diagram of the material; finally, Sec.~\ref{ssec_analytics} shows that quantum anomalous Hall behaviors are primarily induced by the Fock contribution, allowing us to analytically predict the range of moir\'e lattice constants yielding quantum anomalous Hall behaviors for the different rhombohedral $N$-layer graphene stacks.

\subsection{Model} \label{ssec_modelpentalyaer}

\begin{figure}
\centering
\includegraphics[width=0.75\columnwidth]{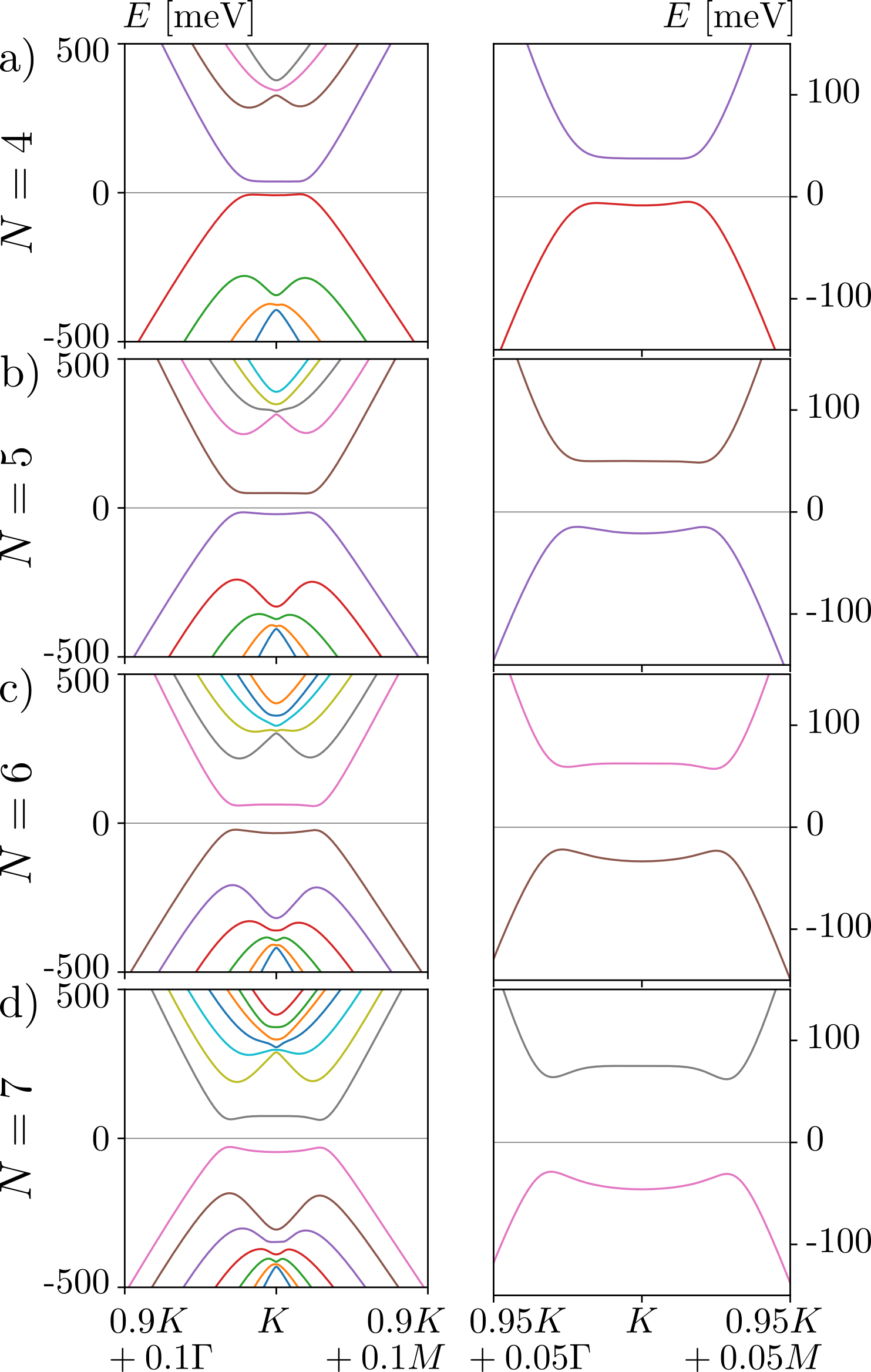}
\caption{Band structure of rhombohedral $N$-layer graphene along the $\Gamma-K-M$ line near the $K$ point, computed with the tight-binding model from Refs.~\cite{zhang2010band,herzog2023moir} with a displacement field corresponding to $E_z = \SI{25}{\milli\electronvolt}$ and a uniform potential on the bottom layer $V_0 = \SI{29}{\milli\electronvolt}$ derived in Ref.~\cite{moon2014electronic}. 
The subfigures (a-d) correspond to different numbers of layers $N$ as indicated on the left. They feature a zoomed out view of the band structure (left) where all bands of the tight binding model are visible along the line $(0.9K+0.1\Gamma) \to K \to (0.9K+0.1M)$, and a close up near charge neutrality (right) along the path $(0.95K+0.05\Gamma) \to K \to (0.95K+0.05M)$. 
}
\label{fig_bandstructures}
\end{figure}

To describe rhombohedral $N$-layer graphene, we use the tight-binding model developed in Ref.~\cite{zhang2010band} with the effect of a displacement field included as a layer dependent potential energy $E_\ell = E_z [\ell - \frac{N-1}{2}]$ with $\ell = 0, \cdots , N-1$ denoting the layer index. 
All results presented in this section are obtained using the tight-binding coefficients from Ref.~\cite{herzog2023moir}; the results only minimally change with other choices of tight-binding coefficients~\cite{koshino2010interlayer,chittari2019gate,zhou2021half,kwan2023moir}. 
We also include in the band structure the uniform energy shift experienced by the atoms closest to the nearly aligned hBN due to virtual second-order tunneling processes. 
As derived in Ref.~\cite{moon2014electronic}, this effect can be described by a potential $V_0 P$ with $V_0 = \SI{29}{\milli\electronvolt}$, where $P$ denotes the projection onto the $B$-sublattice of the bottom layer (\textit{i.e.} on the atoms closest to the substrate after inclusion of lattice relaxation effects):
\begin{equation} \label{eq_Projector}
[P]_{(\ell,\tau), (\ell',\tau')} = \delta_{\ell, \ell'}  \delta_{\tau, \tau'} \delta_{\ell,0} \delta_{\tau,B} ,
\end{equation}
with $\tau\in\{A,B\}$ labeling the sublattice index. 
In Fig.~\ref{fig_bandstructures}, we show the band structure of $N$-layer rhombohedral graphene around the $K$-point, where the gap at charge neutrality is smallest, with $E_z = \SI{25}{\milli\electronvolt}$ and $N=4,\cdots,7$.

\begin{figure}
\centering
\includegraphics[width=0.8\columnwidth]{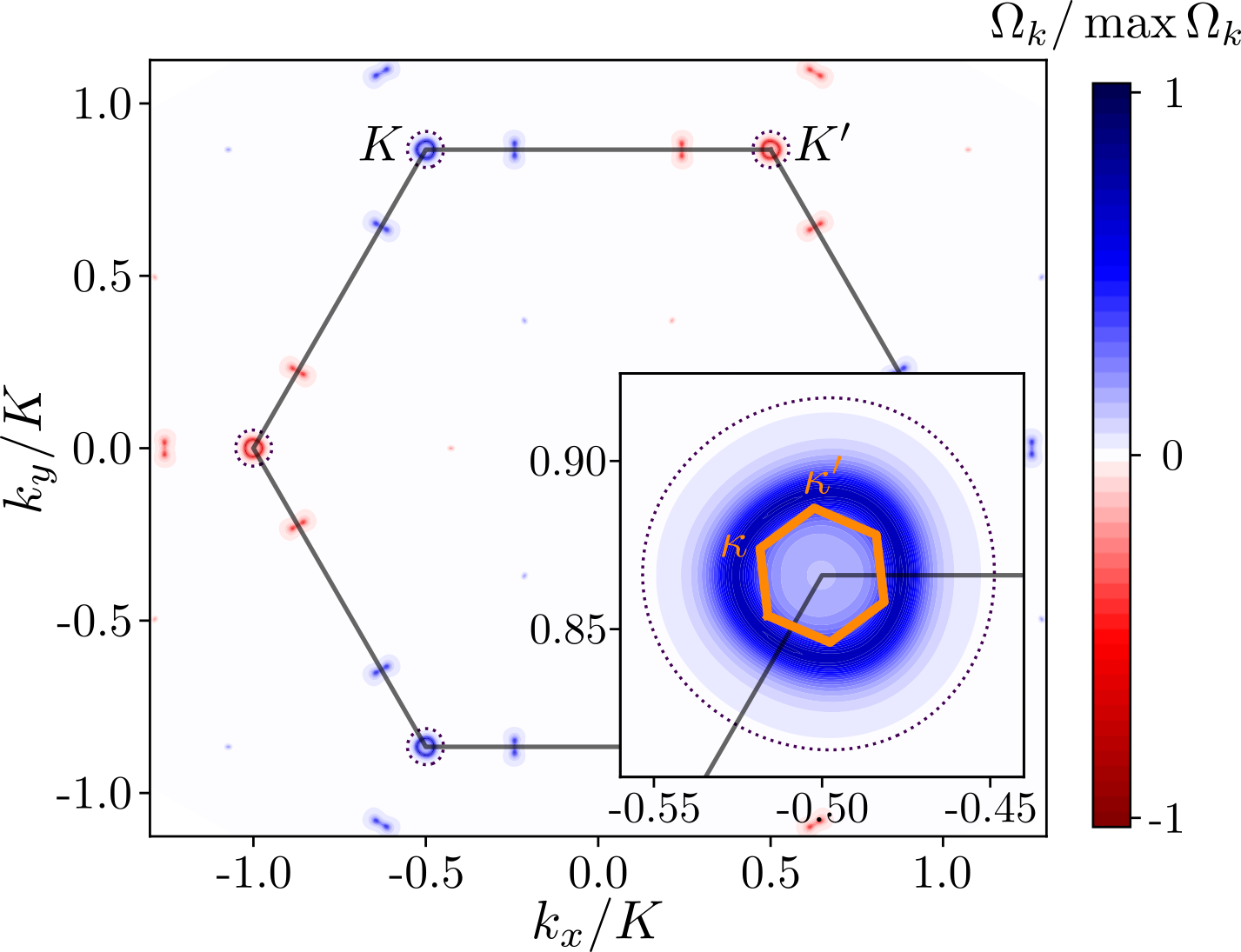}
\caption{The single-band formalism of Secs.~\ref{sec_noninteracting} and~\ref{sec_HartreeFock} only applies in the small regions in momentum space where the lowest conduction band of rhombohedral $N$-layer graphene exhibits an indirect gap to higher bands.
These regions are encircled by a dashed line for the specific case of $N=5$ in the presence of a displacement field $E_z = \SI{25}{\milli\electronvolt}$.  
The regions only span a small fraction of the graphene stack's BZ (black line) near the $K$ and $K'$ points, but nevertheless contain most of the Berry curvature (color scale, also computed for $N=5$ and $E_z = \SI{25}{\milli\electronvolt}$), which is concentrated in rings around those points -- see inset. 
The lattice mismatch and twist of the substrate fixes the position of the corners $\kappa$ and $\kappa'$ of the folded rBZ, represented by a thick orange line for a twist of $0.7^\circ$ and lattice mismatch corresponding to graphene on hBN.
}
\label{fig_locatesingleparticleregime}
\end{figure}

One requirement for the band projected formalism of the previous sections to apply is that the conduction band of the $N$-layer graphene stack be isolated throughout the reduced Brillouin zone set by the superlattice potential. 
As can be seen in Fig.~\ref{fig_bandstructures}, an indirect band gap between the lowest and higher conduction bands only exists in a narrow region near the $K$-point. 
This is the small area on which we now focus to apply the results derived in Secs.~\ref{sec_noninteracting} and~\ref{sec_HartreeFock}. 
To better visualize this narrow region of interest, we represent the full Brillouin zone of rhombohedral penta-layer ($N=5$) graphene in Fig.~\ref{fig_locatesingleparticleregime} and encircle with dashed lines the points where its lowest conduction band remains below the minimum of the second conduction band for $E_z = \SI{25}{\milli\electronvolt}$. 
The inset of Fig.~\ref{fig_locatesingleparticleregime} shows that this single-band regime only applies to momenta $k=K+q$ distant from the graphene's stack $K$ point by at most $|q/K| \simeq 0.04$. 
Similar boundaries are obtained for $N=4 , \cdots , 7$ so long as $E_z \leq \SI{50}{\milli\electronvolt}$. 
While the identified single-band regime only spans a tiny fraction of the $N$-layer original Brillouin zone, it proves to be relevant for the experiments of Refs.~\cite{lu2024fractional,xie2024even}, as we now explain: the nearly-aligned hBN substrate imposes a superlattice potential on the graphene stack whose periodicity and orientation are determined by the lattice mismatch $\epsilon=0.0163$ between graphene and hBN and the relative twist angle $\theta$ between them. 
We determine where the $\kappa$ corner of the superlattice rBZ is located within graphene’s BZ from these factors using $\kappa = K - \frac{1}{1+\epsilon} R(\theta) K$, where $R(\theta)$ denotes the rotation by angle $\theta$~\cite{herzog2023moir}.  
We will later present results as a function of the location of $\kappa$, which is a way of parameterizing the superlattice periodicity and orientation. 
In Ref.~\cite{lu2024fractional}, the twist angle $\theta \simeq 0.7^\circ$ implies a superlattice period $a_{\rm super} \simeq \SI{11.5}{\nano\meter}$ corresponding to corners of the rBZ at a relative distance $|\kappa / K| \simeq 0.02$ from the penta-layer $K$-point. 
The rBZ obtained for these values is depicted with a thick orange line in inset of Fig.~\ref{fig_locatesingleparticleregime}.

Fig.~\ref{fig_locatesingleparticleregime} also shows the Berry curvature of the lowest conduction band of penta-layer graphene when $E_z = \SI{25}{\milli\electronvolt}$, 
which is concentrated in a small ring around the $K$ point, entirely contained within the single-band regime. 
There are additional peaks of Berry curvature induced by trigonal warping along the $K-M$ and $K-\Gamma$ lines far beyond the identified region of interest; hence these do not play a substantial role in our story. 

\subsection{Substrate-induced topology} \label{ssec_pentalayernoint}

In the experiments of Refs.~\cite{lu2024fractional,xie2024even}, the quantized anomalous Hall effect has only been observed for displacement fields polarizing the electronic wavefunctions of the conduction band on the layer located furthest away from the hBN substrate giving rise to the superlattice potential. 
In our notation, this can be modeled with a positive displacement field $E_z>0$ and a long-wavelength potential that only couples to the bottom layer $\ell = 0$~\cite{park2023topological}. 
In fact, the effect of the potential on the $B$ sublattice of the bottom layer alone (\textit{i.e.} on the carbon atoms of the bottom graphene sheet with no direct vertical neighbor in the second layer) was found sufficient to capture the full effect of the hBN substrate on the bottom of the conduction band in the experimentally relevant range of displacement fields~\cite{herzog2023moir}.

We therefore include the effect of the substrate as follows. First, the lattice mismatch and twist of the substrate fix the corners $\kappa$ and $\kappa'$ of the rBZ, represented by a thick orange line in Fig.~\ref{fig_locatesingleparticleregime}. 
Then, the superlattice potential induced by the substrate is introduced as 
\begin{equation}
V_{\rm hBN} (r) = - P \sum_g V_g e^{ig \cdot r} ,  
\end{equation}
with $P$ representing the projection onto the $B$-sublattice of the bottom layer (see Eq.~\ref{eq_Projector}).
This internal structure of the potential does not change the form of the band-projected Hamiltonian appearing in Eq.~\ref{eq_superlatticehamiltonian} nor the derivation presented in Sec.~\ref{sec_noninteracting} provided we use layer and sublattice weighted form factors 
\begin{equation} \label{eq_layerweightedformfactors}
\Lambda_{q',q}^{\rm ls} = \braOket{\chi(q')}{P}{\chi(q)} 
\end{equation}
instead of the bare form factors of the conduction band. 
Finally, we only consider the lowest harmonics of the superlattice, which are equal for the three $\mathcal{C}_3$ related harmonics $g_1 = \kappa+\kappa'$, $g_2 = \kappa-2\kappa'$ and $g_3 = -2\kappa + \kappa'$. 
We adopt the notation introduced in Tab.~\ref{tab_fullsummary} ($n=3$ column)
and write $V_{g_{1,2,3}} = |V| e^{i\theta}$. 
For consistency with the $V_0$ used in the tight-binding calculation, we shall use the values $|V| = \SI{21}{\milli\electronvolt}$ and $\theta = 196.6^\circ$ from Ref.~\cite{moon2014electronic}.

\begin{figure}
\centering
\includegraphics[width=0.85\columnwidth]{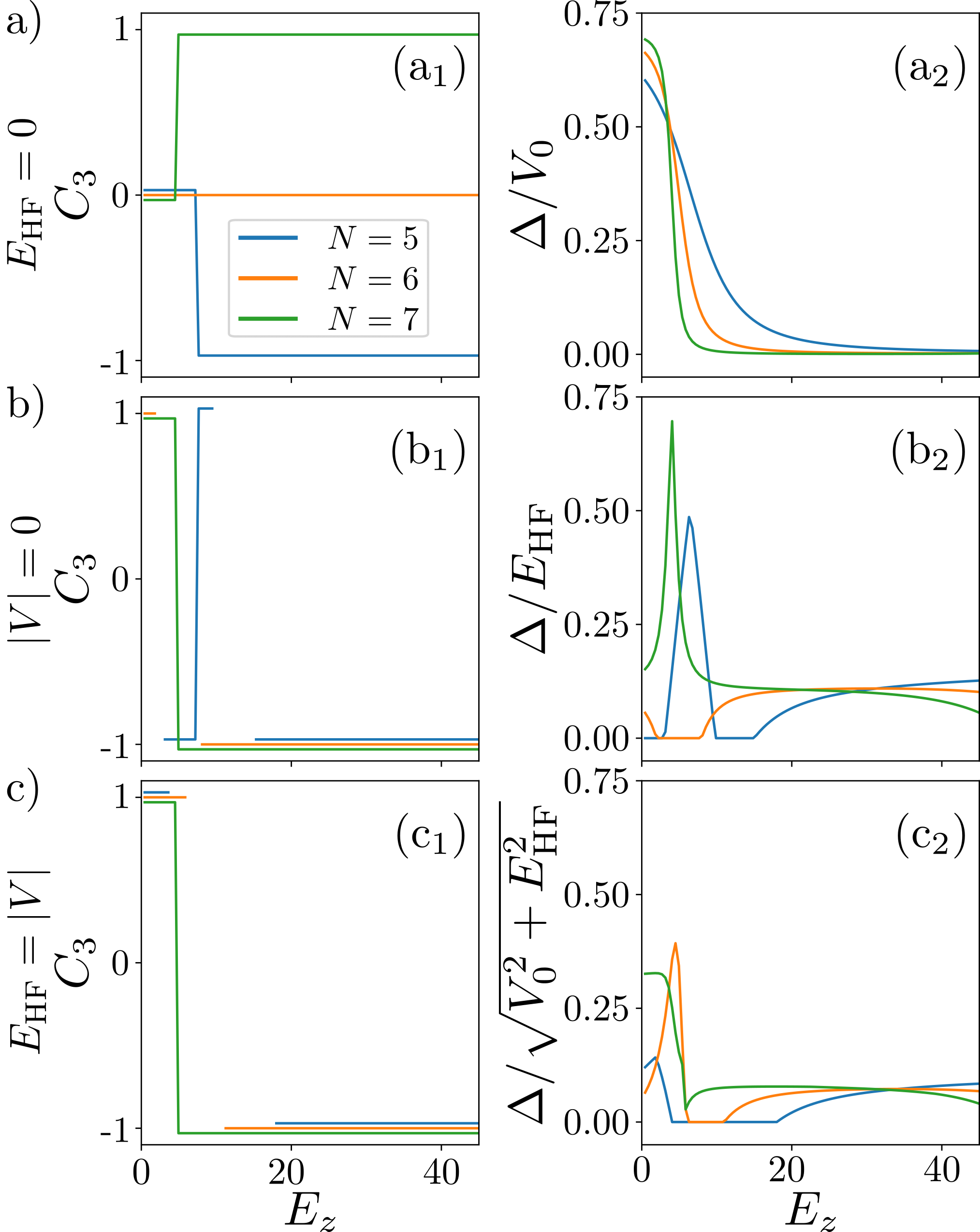}
\caption{Perturbative phase diagrams of rhombohedral $N$-layer graphene as a function of the applied displacement field for different interaction strength $E_{\rm HF} = 0$ (a), in absence of superlattice potential $|V| = 0$  (b), and for $E_{\rm HF} = |V|$ (c). 
The left panels (subscript 1) show the Chern number of the lowest conduction band, obtained with the non-interacting formalism of Sec.~\ref{sec_noninteracting} for (a) and the self-consistent Hartree-Fock calculation of Sec.~\ref{sec_HartreeFock} in (b-c). 
In (b-c), the absence of lines indicates that no perturbative self-consistent solution exists, leading to metallic behavior. 
The right panels (subscript 2) show the gap opened by the superlattice.
}
\label{fig_resultsgraphenestacks}
\end{figure}

To draw the non-interacting topological phase diagram of the $N$-layer graphene stacks, we vary the displacement field $E_z$ in the experimentally relevant range $0-\SI{40}{\milli\electronvolt}$ and use the expression for the Chern number modulo three $C_3$ from Tab.~\ref{tab_fullsummary} (see also App.~\ref{app_combineeigvals}). 
We choose $\kappa$ as illustrated in Fig.~\ref{fig_locatesingleparticleregime}, which corresponds to a $\theta = 0.7^\circ$-twisted hBN substrate. 
Our results, shown in Fig.~\ref{fig_resultsgraphenestacks}a$_1$ for $N=5$, $6$ and $7$, agree with the expected behavior of rhomobohedral graphene in the absence of interactions~\cite{min2008chiral}: we observe a transition from a trivial $C_3=0$ insulator to a topological state with $C_3 = N$ (mod $3$) as $E_z$ increases above a threshold approximately given by $E_z \simeq V_0/(N-1)$~\cite{herzog2023moir} for which a topological gap closing at the rBZ center ($\gamma$) occurs.

To understand the physical relevance of this result, we have also evaluated the non-interacting gap $\Delta = \min (\Delta_\kappa, \Delta_{\kappa'})$, where $\Delta_{q_0}$, the gap opened by the superlattice at the high-symmetry point $q_0$, is determined using Eq.~\ref{eq_energies} to find the energy difference between the lowest two minibands at each high-symmetry point.  
We plot this gap in Fig.~\ref{fig_resultsgraphenestacks}a$_2$, where we observe that $\Delta$ rapidly decays as the displacement field $E_z$ increases. 
This is a consequence of the projector appearing in Eq.~\ref{eq_layerweightedformfactors}. 
Indeed, the displacement field polarizes the conduction band on the layer located furthest away from the one where the superlattice potential acts, which decreases the amplitude of the layer and sublattice weighted form factor $\Lambda_{R_3 \kappa , \kappa}^{\rm ls}$. 
In fact, it is known that, near the $K$-point of the $N$-layer graphene, the weights of conduction band's wavefunctions on the $\ell=0$ layer decrease exponentially fast with $E_z$~\cite{zhang2010band}, which explains the exponential decay of the gap $\Delta_{\kappa}$ in Fig.~\ref{fig_resultsgraphenestacks}a$_2$. 
We emphasize that the topological transitions in Fig.~\ref{fig_resultsgraphenestacks}a$_1$ are associated with a gap closing between the valence and conduction bands at $\gamma$, and therefore do not lead to a vanishing value of $\Delta = \min (\Delta_\kappa, \Delta_{\kappa'})$, which measures the gap to higher conduction bands.

\subsection{Interaction-induced topology} \label{ssec_pentalayerincludeint}

To summarize this non-interacting calculation, we have found that the presence of the nearly aligned hBN substrate yields, for sufficiently large $E_z$,  a topological lowest miniband with Chern number consistent with $C_3 = N$ (mod $3$), whose gap to higher conduction bands decreases exponentially as the displacement field $E_z$ grows due to the polarization of the conduction band in the layer furthest away from the substrate. In particular, this exponential suppression suggests that the non-interacting approximation breaks down for large $E_z$, and that the dominant gap at $\kappa$ is mostly induced by interaction effects -- to which we now turn.

To find the topological character of the lowest conduction band for large displacement fields, we momentarily turn off the superlattice potential ($|V|=0$) and include the effects of interactions within the Hartree-Fock approximation. 
Following the method summarized in Tab.~\ref{tab_summaryhf} and derived in Sec.~\ref{sec_HartreeFock}, we determine the Chern number $C_3$ of the lowest conduction band in the rBZ for which a perturbative self-consistent Hartree-Fock solution exists.

We display the resulting purely-interacting Hartree-Fock phase diagrams in Fig.~\ref{fig_resultsgraphenestacks}b for the same choice of $\kappa$ as in Fig.~\ref{fig_resultsgraphenestacks}a (see inset of Fig.~\ref{fig_locatesingleparticleregime}). 
For all considered values of $E_z$ and $N$, we find at most one $C_3$ value for which a self-consistent Hartree-Fock solution exists using our algorithm. 
When no such $C_3$ is found, our perturbative Hartree-Fock calculation predicts a metallic state. 
These metallic regions, mostly found at intermediate values of the displacement field $E_z \sim 5-\SI{10}{\milli\electronvolt}$, appear in Fig.~\ref{fig_resultsgraphenestacks}b$_1$ as domains with no data points (the Chern number is not defined for gapless states) corresponding to vanishing-gap regions in Fig.~\ref{fig_resultsgraphenestacks}b$_2$.

The main feature of this purely interacting calculation is the presence of a topological insulator with $C_3 = -1$ at large $E_z>\SI{15}{\milli\electronvolt}$ for all $N$ considered in Fig.~\ref{fig_resultsgraphenestacks}. 
This correctly captures the experimental observation of the quantized anomalous Hall effect at integer filling of the moir\'e hBN/rhombohedral graphene systems of Refs.~\cite{lu2023fractional,xie2024even}. 
Finally, we comment that the gap closing at $\gamma$ remains visible in the $N=5$ and $N=7$ interacting phase diagram as a topological transition for which $\Delta$ does not vanish (see discussion above). We observed that the  two excited states above $\kappa$ cross at these transition points, leading to a cusp in the gap depicted at Fig.~\ref{fig_resultsgraphenestacks}b$_2$.

To obtain a complete picture intertwining the non-interacting results of Fig.~\ref{fig_resultsgraphenestacks}a and the Hartree-Fock insights of Fig.~\ref{fig_resultsgraphenestacks}c, we reintroduce the superlattice potential ($|V| > 0$) and use the full-fledged Eq.~\ref{eq_nupHF} to draw the perturbative Hartree-Fock phase diagrams of the rhombohedral $N$-layer graphene stacks as a function of displacement field. 
This requires us to introduce a phenomenological parameter $E_{\rm HF} = \eta v(g_\kappa)$ measuring the strength of the Hartree-Fock potential built-up within the small resonant regions described in Sec.~\ref{ssec_resonantapprox}. 
With $\varepsilon = 6$, we estimate this parameter to be equal to $\frac{E_{\rm HF}}{|V|} \sim \frac{e^2}{4 \pi \varepsilon_0 \varepsilon a_{\rm super} |V|} \simeq 1$ and show the results obtained for this realistic value in Fig.~\ref{fig_resultsgraphenestacks}b. 
For $E_z > \SI{20}{\milli\electronvolt}$, the results almost perfectly agree with the purely interacting case displayed in Fig.~\ref{fig_resultsgraphenestacks}c, while the interacting and non-interacting contribution are similar in magnitude and compete for smaller displacement fields.

We stress that our perturbative calculation does not aim to quantitatively account for all experimental features. Rather, its intent is to rapidly sketch the phase diagram of a particular material using a few material-specific coefficients. In the case of rhombohedral $N$-layer graphene, these coefficents are the form factors $\Lambda$ and layer-sublattice weighted form factors $\Lambda^{\rm ls}$. With only a handful of these coefficients, we are able to qualitatively capture the phases observed in the experiments of Refs.~\cite{lu2024fractional,xie2024even}.

\subsection{Fock term and analytical insights} \label{ssec_analytics}

\begin{figure*}
\centering
\includegraphics[width=\textwidth]{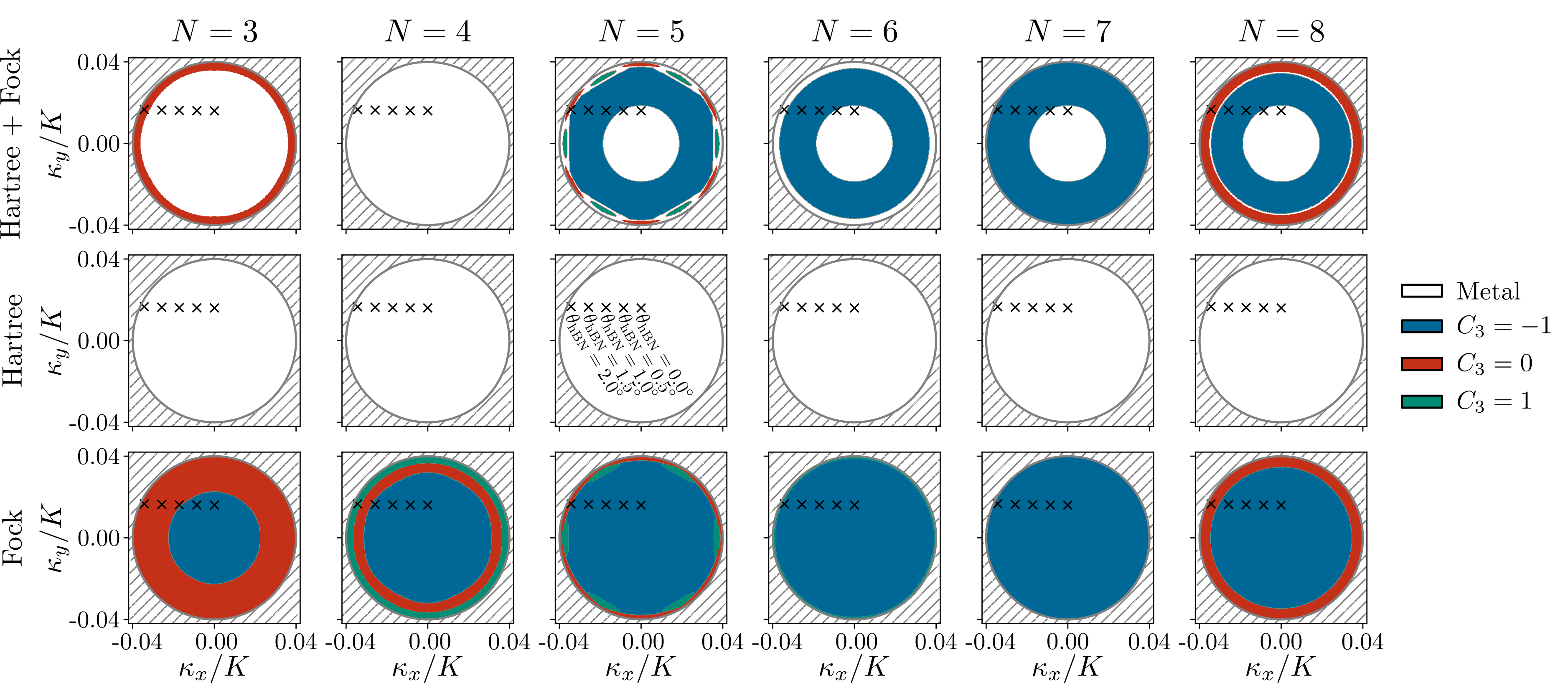}
\caption{Hartree-Fock (top row), Hartree (middle row) and Fock (bottom row) purely interacting phase diagram of rhombohedral $N$-layer graphene for $N=3, \cdots , 8$ (different columns) and $E_z = \SI{30}{\milli\electronvolt}$ as a function of the position of the $\kappa$ corner of the rBZ relative to the graphene stack's $K$ point. The black crosses represent the value of $\kappa$ corresponding to the rBZ corner set by an hBN substrate with a twist angle $\theta = 0^\circ, 0.5^\circ, 1^\circ, 1.5^\circ$ and $2^\circ$, as indicated in the middle row of the $N=5$ column.}
\label{fig_splithartreefock}
\end{figure*}

We now seek additional insight into the non-trivial topology in the experimentally relevant regime of large displacement field $E_z > \SI{25}{\milli\electronvolt}$, where the effect of the superlattice potential is negligible compared to the Hartree-Fock corrections. 
To this aim, we consider the purely interacting case $|V| = 0$ and compare the Chern number of the lowest miniband when we include both Hartree and Fock terms, or only one of the two. 
These Chern numbers, obtained by the algorithm derived in Sec.~\ref{ssec_stronginteractionHF}, are displayed in Fig.~\ref{fig_splithartreefock} for $N=3, \cdots , 8$ and $E_z = \SI{30}{\milli\electronvolt}$ as a function of the position of the $\kappa$ corner of the rBZ respective to the graphene stack's $K$ point. 
We only show results within the single band regime identified in Sec.~\ref{ssec_modelpentalyaer}, \textit{i.e.} for $|\kappa|\leq 0.04 |K|$. For the sake of concreteness, we indicate with black crosses the special points corresponding to the rBZ corner $\kappa$ set by an hBN substrate with a twist angle $\theta = 0^\circ, 0.5^\circ, 1^\circ, 1.5^\circ$ and $2^\circ$.

The first line of Fig.~\ref{fig_splithartreefock} corresponds to the full Hartree-Fock results, and efficiently predicts for which values of $N$ the rhombohedral stack exhibits a topological band under large bias. 
All rhombohedral stacks feature a metallic phase for small mismatch $|\kappa/K|$, where no self-consistent Hartree-Fock solutions can be found, and insulating phases for larger $|\kappa/K|$. 
The exact value of $|\kappa/K|$ at which the transition between the metallic and insulating phases occur, and the Chern number of the insulating phase, vary depending on the number of layers $N$. 
For instance, the transition occurs for $\kappa$ outside of the single-band regime for $N=4$. 
There is, nevertheless, a striking similarity between all mutlilayers with $N\geq 5$: the first insulating state appears for $|\kappa/K| \simeq 0.02$ and has Chern number $C_3 = -1$. 
In particular, all $N\geq 5$ rhombohedral stacks should feature quantum anomalous Hall behavior under large biases when the rBZ is set by a hBN substrate with a twist angle $\simeq 1^\circ$. 
Our results may also explain why, to this date, quantum anomalous Hall behaviors were observed in penta- and hexa-layer graphene but not in quadri-layer graphene.

Comparing the full Hartree-Fock solutions to those obtained by only considering the Hartree (second line of Fig.~\ref{fig_splithartreefock}) and Fock (third line of Fig.~\ref{fig_splithartreefock}) corrections, we observe that the Hartree term always favors a metallic phase while the Fock term gives rise to self-consistent insulating states, and in particular to those with non-zero Chern number. 
From Fig.~\ref{fig_splithartreefock}, it hence appears that the Hartree contribution dominates for small $\kappa$, leading to a metallic state in the full-fledged solution, while the Fock terms take over for $|\kappa / K| \geq 0.02$.

To qualitatively understand the competition between the Hartree and Fock terms, let us simplify the equations in Sec.~\ref{ssec_stronginteractionHF} under the assumptions that ($i$) the particles interact through a short-range interacting potential such that $v(\kappa) \simeq v(g_\kappa)$; ($ii$) the band structure of the $N$-layer graphene can be reduced to the two-band model with continuous rotation symmetry of Ref.~\cite{min2008chiral}, 
\begin{equation}
\mathcal{H}_{\rm eff} (k) = \begin{bmatrix} \Delta & \frac{v_F^N}{t_1^{N-1}} (k_x - ik_y)^N \\  \frac{v_F^N}{t_1^{N-1}} (k_x + ik_y)^N  & - \Delta \end{bmatrix},
\end{equation}
parameterized by the displacement field $\Delta = E_z (N-1)/2$, graphene's Fermi velocity $v_F = \SI{550}{\milli\electronvolt.\nano\meter}$,  and the tunneling $t_1 = \SI{360}{\milli\electronvolt}$ between nearest inter-layer neighbors~\cite{herzog2023moir}; and ($iii$) the ratio of form factors $\frac{\Lambda_{k,q} \Lambda_{q+g,k+g}}{\Lambda_{q+g,q} \Lambda_{k,k+g,k}}$ appearing in the Fock term can be approximated by a phase factor equal to the integrated flux of Berry curvature contained in the $(q\to k \to k+g \to q+g \to q)$ loop discussed below Eq.~\ref{eq_HartreeFock}. 
Under these approximations, we can intuitively understand why the Fock term is negligible for small $\kappa$ and eventually dominates over the Hartree term for $|\kappa/K| \sim 0.02-0.03$, where it drives the system towards an insulator with non-zero Chern number $C_3=-1$.  
To see this, we use approximations $(i-iii)$ to rewrite the $\Gamma$ factors of Eq.~\ref{eq_definitionsofGammas} as 
\begin{equation}  \label{eq_effectivetheoryfock}
\Gamma_{\kappa/\kappa'}^{(3)} \simeq 1 + \omega_3^{C_3} (1 - e^{i\phi_{\rm F}} )  , \quad \phi_{\rm F} = \int_{\mathcal{L}_\kappa} {\rm d}^2 k \, \Omega_k ,
\end{equation} 
with $\mathcal{L}_\kappa$ denoting the region enclosed in momentum-space loop $(\kappa \to \kappa' \to \kappa'+g_\kappa \to \kappa+g_\kappa \to \kappa)$, and where the Berry curvature expanded to leading order in $k$ reads~\cite{herzog2023moir}
\begin{equation}
\Omega_k = \frac{N^2}{2} \frac{v_F^2}{\Delta^2} \left( \frac{v_F |k|}{t_1} \right)^{2(N-1)} . 
\end{equation}
While we do not have an analytic formula for the integrated Berry flux $\phi_{\rm F}$, it is straightforward to integrate $\Omega_k$ in a circle of radius $|\kappa|$. 
Then, to estimate the critical $\kappa$ where the Fock term exceeds the Hartree term,
we write $\phi_{\rm F}$ as the flux threading this circle scaled by a factor $\xi_\kappa \sim 1$ that we have to numerically evaluate,
\begin{equation} \label{eq_fluxberry}
 \phi_{\rm F}  = \frac{\pi \xi_\kappa N^2}{2} \frac{t_1^2}{\Delta^2} \left[ \frac{v_F|\kappa|}{t_1} \right]^{2N} .
\end{equation}
For all the $\kappa$ and $N$ considered, we found $\xi_\kappa$ to be contained between one half and one.

For $\kappa = 0$, the momentum space loop discussed above is reduced to a point such that the Berry flux $\phi_{\rm F} = 0$, resulting in a vanishing the Fock contribution $(1-e^{i\phi_{\rm F}}) = 0$. 
This only leaves a Hartree term in $\Gamma_{\kappa/\kappa'}^{(3)} = 1$, which does not yield any self-consistent solutions (as can be straighforwardly checked with Eq.~\ref{eq_selfconsistentsolutionstronginteraction}). 
This explains the dominance of the Hartree term and the metallic phase at small $\kappa$ for all $N$, see Fig.~\ref{fig_splithartreefock}. 
As $|\kappa|$ increases, the system eventually reaches a point where $\phi_{\rm F} = \pi/3$, for which the equations Eq.~\ref{eq_selfconsistentsolutionstronginteraction} display their first self-consistent solution, which has Chern number $C_3 = -1$ as can be straighforwardly checked from the Eq.~\ref{eq_effectivetheoryfock}. 
Approximating $\xi_\kappa \simeq 0.5$ for the sake of concreteness, the transition occurs for
\begin{equation} \label{eq_matchingcondition}
\frac{|\kappa^*|}{K} \simeq \frac{t_1}{v_F K} \left[ \frac{(N-1) E_z}{\sqrt{3} N t_1 } \right]^{1/N} , 
\end{equation}
which is contained between 0.020 and 0.026 for $N = 5, \cdots , 8$. These values all belong to the $C_3=-1$ insulating phase observed in Fig.~\ref{fig_splithartreefock} for $N\geq 5$.

The exceptional coincidence is that this very specific range, $0.020 < |\kappa^*/K| < 0.026$, exactly matches the size of the rBZ imposed by a hBN substrate twisted by about one degree -- see crosses in Fig.~\ref{fig_splithartreefock}. 
According to our perturbative insights, this coincidence between the hBN-graphene mismatch and the value of $0.6 t_1/(v_F K)$ is the main reason for observing quantum anomalous Hall behaviors in rhombohedral graphene stacks, where the numerical factor 0.6 comes from the term raised to a power $1/N$ in Eq.~\ref{eq_matchingcondition} that lies between 0.5 and 0.7 for $N=5, \cdots , 8$. 

\section{Conclusion and outlook}

We have developed a highly versatile and computationally inexpensive method for predicting the topological properties of materials or heterostructures subjected to a superlattice potential by combining degenerate perturbation theory with the symmetry indicators method. 
Our analysis applies to externally imposed superlattices, for which we provide a systematic rule to find the Chern number of the superlattice-induced miniband from the harmonics of the superlattice potential and a few material-specific coefficients, as well as to interaction-induced potentials, for which we derive an efficient algorithm to determine all Chern numbers compatible with a self-consistent solution to the Hartree-Fock equations. 
Our approach correctly captures the phases observed in rhombohedral graphene multilayers~\cite{lu2024fractional,xie2024even}, 
and, importantly, offers a unique analytical insight as to why the quantum anomalous Hall effect emerges in these systems. 

While our main emphasis here was to apply the perturbative symmetry indicator method to superlattice materials, it can straightforwardly be extended to twisted homobilayers. In particular, in App.~\ref{app_homobilayers}, we show that it accurately reproduces the topological phase diagram of the continuum model for twisted transition metal dichalcogenide homobilayers~\cite{wu2019topological,Devakul_2021} and of twisted bilayer graphene in the chiral limit where the topology of the bands around zero energy is fully characterized by Chern numbers~\cite{Song_2021}.
In the future, we plan to develop a similar framework to obtain additional topological quantum numbers, such as mirror-Chern or $\mathbb{Z}_2$ invariant, for which symmetry indicator formulas are available~\cite{po2020symmetry,bradlyn2017topological,elcoro2021magnetic}.

\section{Acknowledgements}

The Flatiron Institute is a division of the Simons Foundation. 
We thank Y. Zeng for a discussion on a closely related topic that sparked our interest for the problem discussed in this manuscript. 
V.C. thanks N. Regnault for a critical discussion of the results. 
J.C. acknowledges support from the Air Force Office of Scientific Research under Grant No. FA9550-20-1-0260 and from the Alfred P. Sloan Foundation through a Sloan Research Fellowship.

\appendix

\section{Non-interacting Chern number for all $n$} \label{app_combineeigvals}

In this appendix, we combine Eqs.~\ref{eq_nupforq0} and~\ref{eq_formalcombination} to populate the summary given in Tab.~\ref{tab_fullsummary}. For simplicity, we treat the $n=2,3,4,6$ cases separately.

\paragraph{\underline{$n=2$:}}

In the case $n=2$, the rectangular BZ possesses four high-symmetry points with point symmetry group $\mathcal{C}_2$, conventionally named $\gamma$, $x$, $y$ and $m$.
They are graphically defined in the second row of Tab.~\ref{tab_fullsummary}, where we also show the harmonics of the potential $V_{q=x,y,m}$ connecting the high-symmetry points $x$, $y$ and $m$ to their respective $R_2$-rotated images. 
Since $x,y,m$ are not related by any symmetries, $V_{x,y,m}$ are independent degrees of freedom. 

Following Sec.~\ref{ssec_onehighsym_nonint}, the Chern number of the lowest superlattice-induced miniband can be entirely determined from the three harmonics $V_{q=x,y,m}$ and the four matrix elements $\braOket{q}{U_2}{q}$, with $q\in\{\gamma, x, y, m\}$ and $U_2$ the representation of the $R_2$-rotation acting on the Bloch vectors.
Specifically, using Eq.~\ref{eq_Phi2}, each high-symmetry point $q$ defines an independent loop phase
\begin{equation}
\Phi_{q=x,y,m}^{\circlearrowleft} = 2 \arg \left[ V_{g_q} \frac{\braOket{q}{U_2}{q}^*}{\braOket{\gamma}{U_2}{\gamma}^*} \right] . 
\end{equation}
In terms of these variables, 
Eqs.~\ref{eq_nupforq0} and \ref{eq_formalcombination} yield
\begin{equation} \label{eq_ChernNumber2Full}
C_2 = \sum_{q=x,y,m} \left\lfloor \frac{\pi - \Phi_q^{\circlearrowleft}}{2\pi}  \right\rfloor \quad ({\rm mod} \, 2) ,
\end{equation}
giving the phase diagram in the last row of Tab.~\ref{tab_fullsummary}.

In the limit where the period of the superlattice is long, we use Eq.~\ref{eq_berryphaseforoverlap} to simplify the expression of the $\Phi_2^{\circlearrowleft}(q)$ phases. The loop $q \to R_2 q \to R_2^2 q = q$ does not enclose any area (second line of Tab.~\ref{tab_geometricfactor}), which results in $\Phi_2^{\circlearrowleft}(q) = 2 \arg  V_{g_q} $. 
Furthermore, since $p=2$ is even for the relevant high-symmetry points, the potential $V_{g_q}$ is real, such that $\arg  V_{g_q} \in \{ 0 ,\pi \}$ and $\lfloor (\pi - \Phi_2^{\circlearrowleft}(q))/2\pi\rfloor = - \arg  V_{g_q} / \pi$. 
When plugged in Eq.~\ref{eq_ChernNumber2Full}, this gives 
\begin{equation}
C_2 = - \sum_{q=x,y,m} \frac{\arg V_{g_q}}{\pi} = \frac{\arg (V_{g_x} V_{g_y} V_{g_m})}{\pi} \quad ({\rm mod} \, 2)  , 
\end{equation}
using $x = -x \, ({\rm mod} \, 2)$. We can rewrite this as
\begin{equation}
C_2 =\Theta ( - V_{g_x} V_{g_y} V_{g_m}) \quad ({\rm mod} \, 2)  ,
\end{equation}
which is quoted in the fourth line of Tab.~\ref{tab_fullsummary}. 
Thus, in the long-period limit, the lowest superlattice-induced miniband necessarily has non-trivial topology when an odd number of the harmonics $(V_{g_x}, V_{g_y}, V_{g_m})$ are negative.

\paragraph{\underline{$n=3$:}}

In the case $n=3$, the hexagonal BZ possesses three high-symmetry points with point group $\mathcal{C}_3$ named $\gamma$, $\kappa$, and $\kappa'$ and graphically represented in the second line of Tab.~\ref{tab_fullsummary} (third column), together with the harmonics $V_{g_\kappa}$ and $V_{g_{\kappa'}}$ connecting the high-symmetry points $\kappa$ and $\kappa'$ to their respective $R_3$-rotated images. 
Since the superlattice potential is real and $\mathcal{C}_3$-symmetric, these two elements are complex conjugates, $V_{g_{\kappa'}} = V_{g_\kappa}^*$. As a result, the Chern number of the lowest superlattice-induced miniband is completely determined by  $V_{g_\kappa} = |V| e^{i\theta}$ and the three matrix elements $\braOket{q}{U_3}{q}$, with $q\in\{\gamma, \kappa, \kappa'\}$, where we recall that $U_3$ denotes the representation of the $R_3$-rotation acting on the Bloch vectors. Combining these variables into the two independent loop phases (Eq.~\ref{eq_Phi2})
\begin{equation} \label{eq_PhaseForChernNumber3Full}
\Phi_{\kappa / \kappa'}^{\circlearrowleft} = \pm 3 \theta + 3 \arg \left[ \frac{\braOket{\kappa/\kappa'}{U_3}{\kappa / \kappa'}^*}{\braOket{\gamma}{U_3}{\gamma}^*} \right] ,
\end{equation}
with the $+$ sign corresponding to $\kappa$, the Chern number in Eq.~\ref{eq_formalcombination} can be written as
\begin{equation} \label{eq_ChernNumber3Full}
C_3 = \sum_{q=\kappa / \kappa'} \left\lfloor \frac{\pi - \Phi_q^{\circlearrowleft}}{2\pi}  \right\rfloor \quad ({\rm mod} \, 3) ,
\end{equation}
giving the general phase diagram in the last row of Tab.~\ref{tab_fullsummary}.

In the limit where the period of the superlattice is long, we use Eq.~\ref{eq_berryphaseforoverlap} to simplify the expression of the $\Phi_3^{\circlearrowleft}(q)$ phases. The loop $q \to R_3 q \to R_3^2 q \to R_3^3 q = q$ encloses half of the reduced Brillouin zone, as described by the $p=3$ line of Tab.~\ref{tab_geometricfactor}. The loop phases therefore carry an explicit dependence on the Berry flux $\phi_{\rm B}$ of the original material threading the rBZ, which gives the relation 
\begin{equation} \label{eq_longwavelengthC3}
C_3 = \sum_{\epsilon = \pm 1} \left\lfloor \frac{\pi -\frac{1}{2} \phi_{\rm B} + 3 \epsilon \theta}{2\pi}  \right\rfloor  \quad ({\rm mod} \, 3)  , 
\end{equation}
This formula is reproduced in the fourth row of Tab.~\ref{tab_fullsummary}.

\paragraph{\underline{$n=4$:}}

The high-symmetry points and relevant potential harmonics for the $n=4$ case are shown in the second row of Tab.~\ref{tab_fullsummary} (fourth column). 
The main difference compared to the $n=2$ case is that $x$ and $y$ are now related by $\mathcal{C}_4$ symmetry, so that there is a single relevant harmonic $V_{g_x}$ that enters the Chern number calculation. It connects both $x$ to its $R_2$-rotation and $m$ to its $R_4$-rotated image (recall that below Eq.~\ref{eq_energies} we assumed $a_2=0$, which means that no harmonic couples $m$ to $R_2 m$).
Thus, the Chern number is entirely determined by one harmonic $V_{g_x}$ of the superlattice potential and three matrix elements, or two independent loop phases
\begin{equation}  \begin{split}
\Phi_x^{\circlearrowleft} & = 2 \arg \left[ V_{g_x} \frac{\braOket{x}{U_4^2}{x}^*}{\braOket{\gamma}{U_4^2}{\gamma}^*} \right] , \\ \Phi_m^{\circlearrowleft}  & = 4 \arg \left[ V_{g_x} \frac{\braOket{m}{U_4}{m}^*}{\braOket{\gamma}{U_4}{\gamma}^*} \right] ,
\end{split} \end{equation}
following Eq.~\ref{eq_Phi2}, with $U_4$ the representation of the $R_4$-rotation that acts on the Bloch vectors. In terms of these variables, combining Eqs.~\ref{eq_nupforq0} and \ref{eq_formalcombination} yields
\begin{equation} \label{eq_ChernNumber4Full}
C_4 = 2 \left\lfloor \frac{\pi - \Phi_2^{\circlearrowleft}(x)}{2\pi}  \right\rfloor + \left\lfloor \frac{\pi - \Phi_4^{\circlearrowleft}(m)}{2\pi}  \right\rfloor \quad ({\rm mod} \, 4) ,
\end{equation}
from which the phase diagram in the last row of Tab.~\ref{tab_fullsummary} is derived. The $n=4$ case is the first for which $n$ is not prime, allowing some high-symmetry points (here, $x$) to have a smaller point group symmetry than $\gamma$. This difference is the reason for the factor of two in the first term of the right hand side of Eq.~\ref{eq_ChernNumber4Full}.

In the limit where the period of the superlattice is long, we use Eq.~\ref{eq_berryphaseforoverlap} to simplify the expression of the loop phases. Only the loop connecting $m \to R_4 m \to R_4^2 m \to R_4^4 m\to m$ encloses the full rBZ and receives a Berry flux contribution of $\phi_{\rm B}$. Furthermore, the potential $V_{g_x}$ is real because $n=4$ is even, which further simplifies the loop phases using Eq.~\ref{eq_realpotentialC2}. Altogether, this leads to the long-period formula
\begin{equation}
C_4 = \left\lfloor \frac{\pi - \phi_{\rm B}}{2\pi}  \right\rfloor  \quad ({\rm mod} \, 4)  ,
\end{equation}
quoted in the fourth line of Tab.~\ref{tab_fullsummary}. Note that the sign of $V_{g_x}$ has disappeared in the long-period limit for $n=4$. The physical origin of this observation is that for a square lattice, flipping the sign of $V_{g_x}$ is equivalent to shifting the origin of the system by half of either of the two Bravais lattice vectors of the superlattice potential. The sign of $V_{g_x}$ is therefore purely conventional and does not influence the physics of the minibands.

\paragraph{\underline{$n=6$:}}

We finally turn to $n=6$, for which the hexagonal Brillouin zone possesses three high-symmetry points $\gamma$, $\kappa$ and $m$  defined in the second line of Tab.~\ref{tab_fullsummary} (last column). 
The only relevant harmonic $V_{g_\kappa}=V_{g_m}$ (see Tab.~\ref{tab_fullsummary}) is real because $n=6$ is even. As a result, the Chern number of the lowest superlattice-induced miniband only is completely determined by a single real harmonics $V_{g_\kappa}$ and by three matrix elements combined into the two independent loop phases
\begin{equation} \begin{split}
\Phi_\kappa^{\circlearrowleft} & = 3\arg \left[ V_{g_\kappa} \frac{\braOket{\kappa}{U_6^2}{\kappa}^*}{\braOket{\gamma}{U_6^2}{\gamma}^*} \right] , \\
\Phi_m^{\circlearrowleft} & = 2 \arg \left[ V_{g_\kappa} \frac{\braOket{m}{U_6^3}{m}^*}{\braOket{\gamma}{U_6^3}{\gamma}^*} \right] , 
\end{split} \end{equation}
with $U_6$ the representation of the $R_6$-rotation acting on the Bloch vectors. In terms of these variables, 
\begin{equation}  \label{eq_ChernNumber6Full}
C_6 = 3 \left\lfloor \frac{\pi - \Phi_m^{\circlearrowleft}}{2\pi}  \right\rfloor + 2 \left\lfloor \frac{\pi - \Phi_\kappa^{\circlearrowleft}}{2\pi}  \right\rfloor \;\, ({\rm mod} \, 6) ,\end{equation}
which yields the phase diagram in the last row of Tab.~\ref{tab_fullsummary}.

In the limit where the period of the superlattice is long, we use Eq.~\ref{eq_berryphaseforoverlap} to simplify the expression of loop phases. 
Only the loop defined by $\kappa$ and its $R_3$ rotations encloses a non-zero area and receives a contribution from the Berry curvature of the underlying material. We also use Eq.~\ref{eq_realpotentialC2} to simplify the loop phases because $V_{g_\kappa}$ is real. After some algebra, we obtain the long-period formula
\begin{equation}
C_6 = \Theta(-V_{g_\kappa}) + 2 \left\lfloor \frac{\pi \Theta(V_{g_\kappa}) - \frac{1}{2} \phi_{\rm B}}{2\pi}  \right\rfloor  \quad ({\rm mod} \, 6)  ,
\end{equation}
given in the fourth line of Tab.~\ref{tab_fullsummary}.

\section{Hartree-Fock potential in the resonant approximation} \label{app_getthegamma}

In this appendix, we derive the expression for the Hartree-Fock potential at high-symmetry points given in Eq.~\ref{eq_simplifiedHFintermsofGamma}, and give explicit expressions for the auxiliary variables $\Gamma_{q}^{(n)}$ appearing in this equation for all $n=2,3,4,6$ and all high-symmetry points $q$. 

Let us start by considering the $n=2$ case. We pick a high-symmetry point of the rBZ $q \in \{x,y,m \}$ and define $g_q = R_2 q - q$ as the harmonic appearing in Eq.~\ref{eq_nupHF}. 
From the figure depicting the high-symmetry points in Tab.~\ref{tab_fullsummary}, we can see that the only resonant order parameter $\rho_{k,g_q}$ is the one with $k=q$. 
The latter does not appear in the Fock term (see definition of the primed integral below Eq.~\ref{eq_HartreeFock}). Thus, the only non-zero contribution to $V_{q , g_q}^{\rm HF}$ comes from the Hartree term. 
Using Eq.~\ref{eq_selfconsistentrho}, this contribution reads 
\begin{equation}
V_{q , g_q}^{\rm HF} = \frac{\eta \, v(g_q) }{2} \omega_2^{-\nu_2 (q)} \Lambda_{q, R_2 q} , 
\end{equation}
consistent with the general form given in Eq.~\ref{eq_simplifiedHFintermsofGamma} with 
\begin{equation}
\label{eq_definitionsofGammas}
\Gamma_{q}^{(2)} = 1 , \quad q \in \{x,y,m \} .
\end{equation}

For $n=3$ and for each of the $q \in \{\kappa, \kappa'\}$, we define $g_q = R_3 q - q$ as in Tab.~\ref{tab_fullsummary}. 
This time, there are two resonant order parameters $\rho_{k,g_q}$: the first is $k=q$ and only appears in the Hartree term, while the second is $k=\kappa'$ (resp. $k=R_3^{-1} \kappa$) for $q=\kappa$ (resp. $q=\kappa'$) and appears in both the Hartree and Fock integrals. 
The first contributes a term equal to unity to $\Gamma_{q}^{(3)}$, as in the $n=2$ case. 
The second leads to the term proportional to $\omega_3^{C_3}$, as can be seen from the relation $\nu_3(\kappa) + \nu_3(\kappa') = C_3$ (Eq.~\ref{eq_formalcombination}) and the symmetry eigenvalues appearing in the analytic expression of the order parameter (Eq.~\ref{eq_selfconsistentrho}). 
Using the $\mathcal{C}_3$-symmetry of the form factors and the continuous rotation symmetry of the interaction kernel $v(q)$, these $k\neq q$ contributions lead to final result
\begin{widetext}
\begin{equation} \begin{split}
\Gamma_{\kappa}^{(3)} &= 1 + \omega_3^{C_3} \frac{\Lambda_{\kappa', \kappa'+g_\kappa}}{\Lambda_{\kappa, \kappa+g_\kappa}}  \left[ 1 - \frac{v(\kappa'-\kappa)}{v(g_\kappa)} \frac{\Lambda_{\kappa',\kappa} \Lambda_{\kappa+g_\kappa,\kappa'+g_\kappa}}{\Lambda_{\kappa+g_\kappa,\kappa} \Lambda_{\kappa',\kappa'+g_\kappa}} \right]  , \\
\Gamma_{\kappa'}^{(3)} &= 1 + \omega_3^{C_3} \frac{\Lambda_{\kappa, \kappa+g_\kappa-g_{\kappa'}}}{\Lambda_{\kappa', \kappa'+g_{\kappa'}}} \left[ 1 - \frac{v(\kappa'-\kappa)}{v(g_\kappa)} \frac{\Lambda_{\kappa+g_\kappa-g_{\kappa'},\kappa'} \Lambda_{\kappa'+g_{\kappa'},\kappa+g_\kappa}}{\Lambda_{\kappa'+g_{\kappa'},\kappa'} \Lambda_{\kappa+g_\kappa-g_{\kappa'},\kappa+g_\kappa}} \right] . 
\end{split} \end{equation}

The calculation in the $n=4$ case proceeds in a similar manner using the relation $C_4 = \nu_4(m) + 2 \nu_2(x)$ and the fact that $2 x = -2x \, ({\rm mod} \, 4)$ to get 
\begin{equation} \begin{split}
\Gamma_{m}^{(4)} & =  1  + 2 \omega_4^{C_4} \frac{\Lambda_{x, x+g_x}}{\Lambda_{m, m+g_x}}  \left[ 1 - \frac{v(x-m)}{v(g_x)} \frac{\Lambda_{x,m} \Lambda_{m+g_x,x+g_x}}{\Lambda_{m+g_x,m} \Lambda_{x,x+g_x}} \right] , \\
\Gamma_{x}^{(4)} & = 1  + \frac{1}{2} \omega_4^{-C_4} \frac{\Lambda_{m, m+g_x}}{\Lambda_{x, x+g_x}}  \left[ 1 - \frac{v(x-m)}{v(g_x)} \frac{\Lambda_{m,x} \Lambda_{x+g_x,m+g_x}}{\Lambda_{x+g_x,x} \Lambda_{m,m+g_x}} \right] .
\end{split} \end{equation}
The main difference with the $n=3$ result is that the Chern number $C_4$ here appears with opposite signs in the power of $\omega_4$ for $\Gamma_{m}^{(4)}$ and $\Gamma_{x}^{(4)}$. 
Finally, repeating the calculation for $n=6$ using $C_6 = 2\nu_3(\kappa) - 3 \nu_2(m)$ leads to
\begin{equation} \begin{split}
\Gamma_{\kappa}^{(6)} & = 1  + \frac{3}{2} \omega_6^{C_6} \frac{\Lambda_{m, m+g_\kappa}}{\Lambda_{\kappa, \kappa+g_\kappa}}  \left[ 1 - \frac{v(m-\kappa)}{v(g_\kappa)} \frac{\Lambda_{m,\kappa} \Lambda_{\kappa+g_\kappa,m+g_\kappa}}{\Lambda_{\kappa+g_\kappa,\kappa} \Lambda_{m,m+g_\kappa}} \right]  , \\
\Gamma_{m}^{(6)} & = 1  + \frac{2}{3} \omega_6^{-C_6} \frac{\Lambda_{\kappa, \kappa+g_\kappa}}{\Lambda_{m, m+g_\kappa}}  \left[ 1 - \frac{v(m-\kappa)}{v(g_\kappa)} \frac{\Lambda_{\kappa,m} \Lambda_{m+g_\kappa,\kappa+g_\kappa}}{\Lambda_{m+g_\kappa,m} \Lambda_{\kappa,\kappa+g_\kappa}} \right]  . 
\end{split} \end{equation}
\end{widetext}

\section{Extension to homobilayers} \label{app_homobilayers}

In this appendix, we extend our perturbative calculation of the Chern number to twisted transition metal dichalcogenide (TMD) homobilayers and twisted bilayer graphene (TBG). 
Because our work only discusses symmetry indicators for the Chern number, we restrict ourselves to the chiral limit of TBG~\cite{tarnopolsky_origin2019} where the topology of the flat bands near charge neutrality is characterized by a sublattice resolved Chern number~\cite{Song_2021,crepel_2024topologicallyprotectedflatnesschiral}. 
The main difference between the superlattice case discussed in the main text and the homobilayers discussed here is a doubling of the number of degrees of freedom. 
In particular, the bare dispersion in absence of moir\'e potential now displays two degenerate minima corresponding to the two layers.

\subsection{Twisted transition metal dichalcogenides}     

\begin{figure}
\centering
\includegraphics[width=\columnwidth]{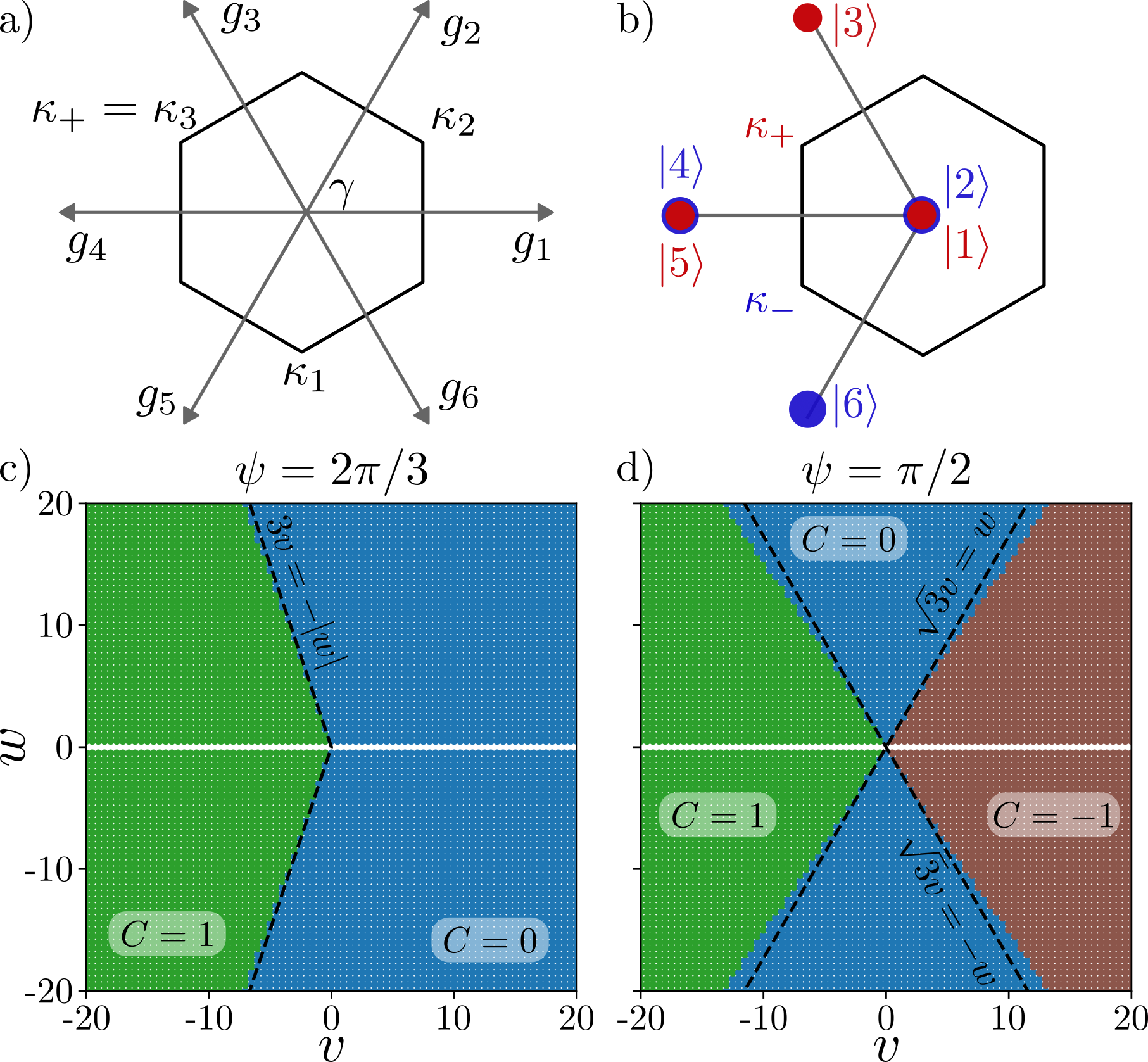}
\caption{
a) Moir\'e BZ indicating the reciprocal lattice vectors $g_{j=1,\cdots,6}$ and the zone corners $\kappa_{j=1,2,3}$. 
b) The single-particle states $\ket{j=1,\cdots ,6}$, obtained by $C_3$ rotation of the state at $\gamma$ in the top (red) and bottom (blue) layers, are degenerate with those at the $\gamma$ point in the absence of moir\'e potential ($v=w=0$).
(c-d) Chern number of the lowest energy band of the twisted TMD model (Eq.~\ref{eq_TMDmodel}) for $\psi = 2\pi/3$ (c) and $\psi = \pi/2$ (d) as a function of the material-dependent parameters $v$ and $w$. Colors show the Chern number obtained by exactly solving the continuum model, while dashed black lines indicate phase boundaries predicted by our perturbative symmetry indicator method. Excellent agreement is found throughout the phase diagram in (c) and for $(v,w) \lesssim 5$ in (d). 
}
\label{fig_bilayers}
\end{figure}

Up to a twist-angle dependent energy scale, the continuum model for the moir\'e bands of a twisted TMD arising from the $K$ valley of the monolayers takes the form~\cite{wu2019topological,Devakul_2021}
\begin{equation} \label{eq_TMDmodel}
H_{\rm TMD} (v, w, \psi)= H_0 (k) + H_m (r) , 
\end{equation}
where
\begin{align} \label{eq_TMDmodelspecifics}
H_0 (k) &= (k - \kappa_{\mu_z})^2 ,  \\
H_m (r) &= 2 v \sum_{j=1,3,5} \cos(g_j\cdot r + \mu_z \psi)  + w \left[ \mu_+ T(r)  + h.c. \right] , \notag \\ 
T(r) & = e^{-i \kappa_+ \cdot r} ( e^{-i \kappa_1 \cdot r} + e^{-i \kappa_2 \cdot r} + e^{-i \kappa_3 \cdot r}  ) e^{i \kappa_- \cdot r} , \notag
\end{align}
with $\vec\mu$ denoting layer Pauli matrices, and where $(v, w, \psi)$ are material-dependent parameters usually extracted from \textit{ab-initio} calculations. 
We have also introduced $g_{j=1,\cdots,6}$, the moir\'e reciprocal lattice vector of smallest norm making an angle $(j-1)\pi/3$ with the positive $k_x$-axis, and $\kappa_{j=1,2,3}$, the three equivalent corners of the Brillouin zone making an angle $2(j-1)\pi/3$ with the negative $k_y$ axis, as shown in Fig.~\ref{fig_bilayers}a. 
Our convention is that the minimum of the dispersion relation of the top and bottom layer lie at the $\kappa_+ = \kappa_3$ and the $\kappa_- = -\kappa_2$ points, respectively. 
The interlayer tunneling $T(r)$ in Eq.~\ref{eq_TMDmodelspecifics} is written to highlight the $C_3$ symmetry of the model and that $T(r)$ moves one electron from the bottom layer with dispersion minimum at $\kappa_-$ to the top layer with the dispersion minimum at $\kappa_+$.
The moir\'e bands from the $K'$ valley are obtained by time-reversal conjugation.

Strong Ising spin-orbit coupling in the TMD monolayers induces spin-valley locking at low energy. 
As a result, spin plays a transparent role in the band topology~\cite{zhang_spin2021,crepel_bridging2024}, only adding a global minus sign to the formula in Eq.~\ref{eq_ChernFromSym}, which we discard for the sake of notation consistency. The geometry of the bilayer requires the symmetry indicator with $n=3$.

As in the main text, we consider the perturbative effect of the moir\'e Hamiltonian $H_m$ on the bare material dispersion $H_0$ to determine the $C_3$ eigenvalues of the lowest energy band at high-symmetry points of the moir\'e Brillouin zone. The main difference compared to a superlattice potential applied to a single layer (discussed in the main text) is that the minima of the dispersion for the two layers are located at different positions, i.e., $\kappa_\pm$, which changes the degenerate subspace to consider in the perturbative calculation outlined in Sec.~\ref{ssec_onehighsym_nonint}. 

Specifically, the band minima of the isolated monolayers are also the band minima of the heterobilayer. Thus, the $C_3$ eigenvalues at the $\kappa_\pm$ points are exactly the $C_3$ eigenvalues of the bare monolayers at their respective $K$ point.
Without loss of generality, this quantity can be taken as unity, \textit{i.e.} $\nu_3(\kappa_\pm)=0$, as the orbital content of the monolayers contributes an overall phase $e^{2i\pi/3}$ to the representation of $C_3$ for all high-symmetry points, which cancels in the symmetry indicator formula Eq.~\ref{eq_ChernFromSym}~\cite{jia2024moire}.

To determine the Chern number of the lowest moir\'e band of Eq.~\ref{eq_TMDmodel},
it remains to determine the $C_3$ eigenvalue of the perturbative ground state at the $\gamma$ point. 
This ground state has already been obtained in Ref.~\cite[App.~D]{Devakul_2021}; we now review its construction. 
Let $\ket{j=1,3,5}$ denote the plane wave in the top layer with momentum $k_j = \kappa_+ + (g_j + g_{j-1})/3$, and $\ket{j=2,4,6}$ the plane wave in the bottom layer with momentum $k_j = \kappa_- + (g_j + g_{j-1})/3$.
As can be seen in Fig.~\ref{fig_bilayers}b, these states transform as $C_3 \ket{j} = \ket{j+2}$ under $2\pi/3$ rotations centered at the minimum of the top ($j$ odd) and bottom ($j$ even) moir\'e BZ -- making them degenerate in energy in absence of moir\'e potentials ($v=w=0$). 
As in the main text, the discrete Fourier transformation 
\begin{equation} \label{eq_bilayerbasisperturbation}
\ket{n} = \frac{1}{\sqrt{6}} \sum_{j=1}^6 e^{2inj\pi/3} \ket{j} 
\end{equation}
perturbatively diagonalizes the moir\'e Hamiltonian $H_m$, yielding the eigen-energies
\begin{equation} \label{eq_perturbativetmd}
\varepsilon_n^{\rm TMD} (v,w,\psi) = 2 w \cos (\pi n/3) + 2v \cos (\psi - 2 n \pi/3) ,
\end{equation}
for the state $\ket{n}$ with $C_3$ eigenvalue $e^{2in\pi/3}$. 
Combined with $\nu_3(\kappa_\pm)=0$, the symmetry indicator formula Eq.~\ref{eq_ChernFromSym} dictates that the lowest moir\'e band is topologically trivial if the minimum $\varepsilon_n$ is given by $n \in \{0,3\}$; has Chern number 1 modulo 3 if $n \in \{1,4\}$; and Chern number $-1$ modulo 3 if $n \in \{2,5\}$.

To demonstrate the predictive power of this method, let us take $\psi = 2\pi/3$~\cite{crepel2024chiral} and compare the Chern number obtained from the perturbative symmetry indicator result to that obtained by solving the full continuum model in Eq.~\ref{eq_TMDmodel}. For this choice of angle $\psi$, Eq.~\ref{eq_perturbativetmd} gives a Chern number equal to 1 when $3 v < -|w|$ and equal to zero otherwise. This perfectly matches the results of the full continuum model, as shown in Fig.~\ref{fig_bilayers}c. This agreement demonstrates the accuracy of our perturbative symmetry indicator method for the twisted TMD bilayers. In  Fig.~\ref{fig_bilayers}d, we show the results for $\psi = \pi/2$, for which Eq.~\ref{eq_perturbativetmd} predicts a Chern number equal to 0 when $|w|>\sqrt{3} |v|$, and away from this region gives a Chern number $1$ when $\sqrt{3} v < -|w|$ and to $-1$ when $\sqrt{3} v > |w|$. Again, the results of the full continuum model and of our perturbative expansion agree very well for small values of $v$ and $w$; they deviate slightly as these two coefficients become very large, corresponding to the small twist angle limit.

\subsection{Chiral twisted bilayer graphene}

To apply our results to TBG, we consider the chiral limit, where the same-orbital inter-layer hopping is discarded~\cite{tarnopolsky_origin2019}, resulting in a purely off-diagonal Hamiltonian in the sublattice degree of freedom and sublattice-polarized flat bands at the magic angle~\cite{Song_2021,crepel_2024topologicallyprotectedflatnesschiral}. 
One way to compute the Chern number of these bands is to square the chiral TBG Hamiltonian, which, owing to the off-diagonal structure of the Hamiltonian, decouples the $A$ and $B$ sublattices as 
\begin{equation}
H_{\rm TBG}^2 (\alpha) = \begin{bmatrix} H_A (\alpha) & 0 \\ 0 & H_B (\alpha) \end{bmatrix} . 
\end{equation}
The resulting squared Hamiltonian only depends on $\alpha$, the ratio of inter-layer tunneling to Fermi velocity. The functional form of $H_{A/B}$ takes the form~\cite{crepel2024chiral}
\begin{align}
H_A (\alpha) & = H_{\rm TMD} (|\alpha|^2, \alpha, \psi = 2\pi/3) + [ \tilde{T}_k(r) \mu_+ +h.c. ] , \notag \\ 
\tilde{T}_k(r)  & = \frac{2 i \alpha}{|\kappa_+|^2} \, e^{-i\kappa_+\cdot r} \left( \sum_{j=1,2,3} e^{-i \kappa_j \cdot r} \kappa_j \times k \right) e^{i \kappa_- \cdot r} .
\end{align}
The Chern number of the band from $H_B$ is opposite to that from $H_A$~\cite{Song_2021}, and we only consider $H_A (\alpha)$ from now on. 
Projecting this Hamiltonian in the perturbative basis defined in Eq.~\ref{eq_bilayerbasisperturbation}, we find an additional term in the energy 
\begin{align}
\varepsilon_n^{\rm TBG} & = \varepsilon_n^{\rm TMD} (|\alpha|^2, \alpha, 2\pi/3) - 2 \sqrt{3} \alpha \sin  (\pi n/3) \\
& =  2 |\alpha|^2 \cos [ 2 (n-1) \pi / 3] + 4 \alpha \cos[ \pi (n+1) / 3]  . \notag
\end{align}
This energy is minimized by $n=2$ when $\alpha>0$ and $n=5$ when $\alpha<0$, both resulting in a $C_3$ eigenvalue of $\nu_3^{A}(\gamma) = -1$. 
Reproducing the equation for the other sublattice $H_B(\alpha)$ yields an opposite value $\nu_3^{B}(\gamma) = 1$, as required by the symmetry of the model~\cite{Song_2021}. 
To summarize, for all $\alpha$, our perturbative calculation predicts a sublattice-polarized Chern number of $\pm 1$, opposite for the two sublattices. Remarkably, this matches the result of the Bistritzer-MacDonald model in the chiral limit~\cite{tarnopolsky_origin2019,Song_2021}. The agreement comes because our perturbative method is justified at small $\alpha$, and the Chern number in the chiral limit remains the same for any value of $\alpha$.

\bibliography{Biblio}

\end{document}